\documentclass[12pt]{article}

\usepackage[colorlinks=true,allcolors=crimson]{hyperref}

\usepackage[
    backend=biber,
    citestyle=authoryear-comp,
    bibstyle=nature,
    sorting=nyt,
    sortcites=true,
    natbib=true
]{biblatex}

\usepackage{setspace}
\usepackage{arxiv}
\usepackage{newpxtext,newpxmath}

\usepackage[utf8]{inputenc} 
\usepackage[T1]{fontenc} 
\definecolor{Silver}{rgb}{0.752,0.752,0.752}
\definecolor{crimson}{RGB}{165,28,48}

\usepackage{url}            
\usepackage{booktabs}       
\usepackage{amsfonts}       
\usepackage{nicefrac}       
\usepackage{microtype}      
\usepackage{lipsum}		
\usepackage{graphicx}
\usepackage{doi}

\usepackage{amsthm}
\usepackage{algorithm2e}
\RestyleAlgo{ruled}
\usepackage{color}
\usepackage[toc,page]{appendix}
\usepackage{placeins}
\usepackage{tabularray}
\usepackage[most]{tcolorbox}
\usepackage[dvipsnames]{xcolor}
\usepackage{soul}

\usepackage{tabularx}
\usepackage{array}

\newcolumntype{Y}{>{\centering\arraybackslash}X} %

\colorlet{reviewchange}{black}
\newcommand{\rev}[1]{{\color{reviewchange}#1}}

\usepackage{amsmath,amsfonts,bm}

\def\figref#1{figure~\ref{#1}}
\def\Figref#1{Figure~\ref{#1}}

\def\eqref#1{equation~\ref{#1}}
\def\Eqref#1{Equation~\ref{#1}}

\def\1{\bm{1}}

\def\rva{{\mathbf{a}}}

\def\rmA{{\mathbf{A}}}

\def\va{{\bm{a}}}
\def\vb{{\bm{b}}}

\def\vq{{\bm{q}}}

\def\vv{{\bm{v}}}

\def\vx{{\bm{x}}}
\def\vy{{\bm{y}}}
\def\vz{{\bm{z}}}

\def\mA{{\bm{A}}}

\def\mD{{\bm{D}}}

\def\mG{{\bm{G}}}
\def\mH{{\bm{H}}}
\def\mI{{\bm{I}}}

\def\mK{{\bm{K}}}

\def\mM{{\bm{M}}}

\def\mP{{\bm{P}}}
\def\mQ{{\bm{Q}}}

\def\mV{{\bm{V}}}
\def\mW{{\bm{W}}}
\def\mX{{\bm{X}}}
\def\mY{{\bm{Y}}}
\def\mZ{{\bm{Z}}}

\def\mOmega{{\bm{\Omega}}}

\DeclareMathAlphabet{\mathsfit}{\encodingdefault}{\sfdefault}{m}{sl}
\SetMathAlphabet{\mathsfit}{bold}{\encodingdefault}{\sfdefault}{bx}{n}

\def\gC{{\mathcal{C}}}

\def\gK{{\mathcal{K}}}
\def\gL{{\mathcal{L}}}

\def\gN{{\mathcal{N}}}
\def\gO{{\mathcal{O}}}
\def\gP{{\mathcal{P}}}

\def\gS{{\mathcal{S}}}

\def\sN{{\mathbb{N}}}

\def\sR{{\mathbb{R}}}

\newcommand{\Ls}{\mathcal{L}}

\DeclareMathOperator{\Tr}{Tr}

\newcommand{\undermax}[1]{\underset{#1}{\max}}
\newcommand{\undermin}[1]{\underset{#1}{\min}}

\newtheorem{theorem}{Theorem}
\newtheorem{lemma}{Lemma}

\newtheorem{proposition}{Proposition}[lemma]

\title{Implicit Generative Modeling by Kernel Similarity Matching}

\author{ \href{https://orcid.org/0009-0006-3838-0389}{\includegraphics[scale=0.06]{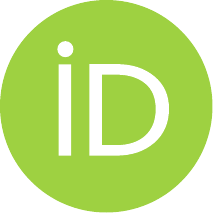}}{\hspace{1mm}Shubham ~Choudhary} \\
	School of Engineering and Applied Sciences\\
    Kempner Institute for the Study of Natural and Artificial Intelligence\\
	Harvard University\\
	Cambridge, MA 02138 \\
	\texttt{shubham\_choudhary@g.harvard.edu} \\
	\And
	\href{https://orcid.org/0000-0003-2001-7515}{\includegraphics[scale=0.06]{orcid.pdf}}{\hspace{1mm}Paul ~Masset}\thanks{PM and DB jointly supervised this work} \\
	Department of Psychology\\
	McGill University\\
	Montréal, QC H3A 1G1 \\
    Mila - Quebec AI Institute 
    \\ Montréal, QC H2S 3H1 \\
	\texttt{paul.masset@mcgill.ca} \\
	\AND
    {\hspace{1mm}Demba ~Ba$^*$} \\
	School of Engineering and Applied Sciences\\
    Kempner Institute for the Study of Natural and Artificial Intelligence\\
	Harvard University\\
	Cambridge, MA 02138 \\
    \texttt{demba@seas.harvard.edu} \\
}

\date{}

\renewcommand{\shorttitle}{}

\hypersetup{
pdftitle={A template for the arxiv style},
pdfsubject={q-bio.NC, q-bio.QM},
pdfauthor={David S.~Hippocampus, Elias D.~Striatum},
pdfkeywords={First keyword, Second keyword, More},
}

\begin{document}
\maketitle

\begin{abstract}
	Understanding how the brain encodes stimuli has been a fundamental problem in computational neuroscience. Insights into this problem have led to the design and development of artificial neural networks that learn representations by incorporating brain-like learning abilities. Recently, learning representations by capturing similarity between input samples has been studied \citep{pehlevan_why_2018} to tackle this problem. This approach, however, has thus far been used only to learn downstream features from an input and has not been studied in the context of a generative paradigm, where one can map the representations back to the input space, incorporating not only bottom-up interactions (stimuli $\to$ latent) but also learning features in a top-down manner (latent $\to$ stimuli). We investigate a kernel similarity matching framework for generative modeling. Starting with a modified sparse coding objective for learning representations proposed in prior work \citep{olshausen1996emergence, tolooshams2021stable}, we demonstrate that representation learning in this context is equivalent to maximizing similarity between the input kernel and a latent kernel. We show that an implicit generative model arises from learning the kernel structure in the latent space and show how the framework can be adapted to learn manifold structures, potentially providing insights as to how task representations can be encoded in the brain. To solve the objective, we propose a novel Alternate Direction Method of Multipliers (ADMM) based algorithm and discuss the interpretation of the optimization process. Finally, we discuss how this representation learning problem can lead towards a biologically plausible architecture to learn the model parameters that ties together representation learning using similarity matching (a bottom-up approach) with predictive coding (a top-down approach).
\end{abstract}



\section{Introduction}
\label{sec:intro}
Learning meaningful representations from data is one of the fundamental goals in engineering applications and scientific discovery. In the field of machine learning, extracting high-quality representations from input data has empowered algorithms to significantly improve their performance on any given task. In neuroscience, the challenge of developing algorithms that can learn representations useful for downstream tasks while respecting the biological constraints of learning is of particular interest, as it may
potentially shed light on the neural basis of behavior. 
Recent approaches in both fields have focused on learning such representations by capturing the relationship between different samples and ensuring that a similar association is maintained within the learned representations. By comparing samples with one another, such unsupervised approaches can help in learning robust representations for downstream tasks. 


In classical machine learning literature, unsupervised methods -- like IsoMap, Multi-dimensional Scaling, t-SNE -- have long been used to learn good representations by measuring some notion of similarity, such as distance, kernel structure etc.  between input samples \citep{cox2000multidimensional, choi2004kernel, van2008visualizing,ghodsi2006dimensionality}. More recently, similarity based representation learning have been explored in the context of deep learning techniques, such as contrastive learning and graph neural nets \citep{chen_simple_2020, zimmermann2021contrastive, sanchez2021gentle}.

\rev{In neuroscience, recent work by \citep{pehlevan_why_2018, luther_kernel_2022} adapt these \textit{similarity based} techniques to develop biologically plausible models for representation learning, providing mechanistic understanding about how neurons in the brain may encode external stimuli. However, while these \textit{similarity-based} measures are effective for feature extraction, they usually do not specify how latent causes generate the sensory input, leaving no principled route for top-down predictions. A \textbf{generative framework} would allow us to achieve this, with higher areas predicting the activity in lower areas through feedback (top-down) mechanisms and feedforward pathways carrying the residual signals, consistent with predictive-coding theories of cortical interactions \citep{keller_predictive_2018, boutin_sparse_2021, rao_predictive_1999}}. For instance, \citep{boutin_sparse_2021} demonstrated in their work that such interactions are useful for explaining important features of neural processing in the visual cortex, such as contour integration, accounting for object perception in the V1 layer even with degraded images, thus shedding light into how we perceive visual stimuli based on expectations and implicit world knowledge, rather than just raw sensory input. Furthermore,  the top-down interactions also play a critical role in regulating information flow from incoming sensory data, ensuring only essential information gets propagated to higher cortical areas, thereby reducing redundancy \citep{keller_predictive_2018}. These studies indicate that integrating top-down interactions in the representation learning process is critical for understanding how the brain encodes and interprets sensory information. 



\begin{figure}[h]
    \hspace{-7mm}
    \includegraphics[scale=0.73]{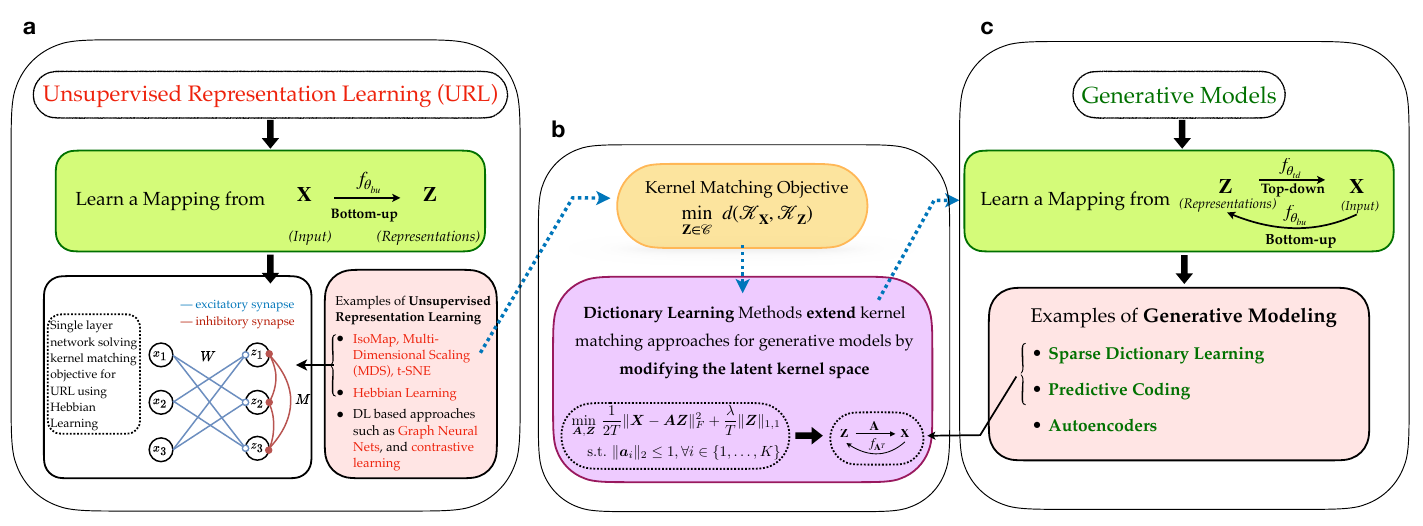}
    \caption{(a) Kernel Similarity Matching (KSM) approaches have been successfully used  previously in the context of unsupervised representation learning which only warrants learning a bottom-up map from the input to the representations. Prior works have utilized this framework in formulating dimensionality reduction techniques such as IsoMap, Multi-Dimensional Scaling (MDS) \citep{ghodsi2006dimensionality} and t-SNE \citep{van2008visualizing}, in addition to developing neurally plausible models for the same.  (c) Generative models, on the other hand involves a bi-directional learning process where the learned representations need to map back to the input space, requiring a top-down map. \rev{Dimensionality reduction methods such as Principal Component Analysis (PCA) or Independent Component Analysis (ICA) have been explored under both categories.} (b) We propose that KSM frameworks can serve as link to extend unsupervised feature learning to generative models. We leverage one class of generative models rooted in dictionary learning methods \citep{kreutz2003dictionary} to extend kernel matching objectives to a generative framework learning objective where the representations not only have to predict the input data but also have to satisfy sparsity constraints. The learning in these cases occur through a combination of both bottom-up (\textit{to estimate latent}) and top-down (\textit{to predict input using the latents}) interaction.}
    \label{fig: overview}
\end{figure}

\begin{figure}[h]
    \centering
    \includegraphics[scale=1.4]{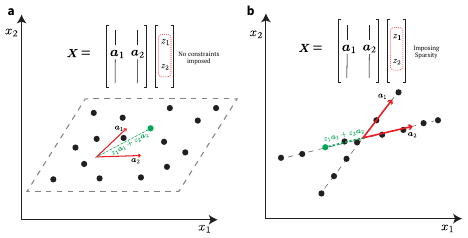}
    \rev{\caption{Nonlinear Manifolds captured by imposing sparsity constraints on the latent representation (b) as opposed to linear subspaces (a) captured without any constraints on the latent representation}}
    \label{fig: non-linearity}
\end{figure}

\rev{Additionally, similarity matching based approaches, thus far, have been adapted to learn linear, typically undercomplete (latent dimension $\le$ input dimension) generative models (e.g. PCA / ICA) \citep{pehlevan_why_2018, bahroun2021normative, minden2018biologically}, or in the case of localized overcomplete manifold-tiling codes (such as Non-negative Similarity Matching (NSM)) \citep{sengupta_manifold-tiling_2018} without an explicit generative map. In this work we extend kernel matching approaches to model \textbf{nonlinear manifolds} by focusing on sparse dictionary learning methods \citep{olshausen_sparse_1997} (\Figref{fig: overview})}. Sparse dictionary learning refers to a class of generative models that have been widely used in the field of signal processing, machine learning and neuroscience to obtain parsimonious representations of the input data, at the same time learning a dictionary of atoms that map the learned representations back to the input space. \rev{In the following work, we show that sparse dictionary learning methods can enable kernel-based approaches can learn \textbf{overcomplete}, constraint-driven representations, with the latent kernel structure providing an implicit generative map with receptive-field like characteristics}. Mathematically, the problem can be defined as:

\begin{equation}
   \label{eq:orig_sparse_coding}
   \begin{aligned}
      \underset{\mA, \mZ}{\min}&\,\, \frac{1}{2T}\|\mX - \mA\mZ\|_F^2 + \frac{\lambda}{T}\|\mZ\|_{1, 1}\quad\text{s.t.}\,\, \|{\va}_i\|_2 \leq 1, \forall i \in \{1, \dots, K\}\\
      \rev{\implies \underset{\mA, \mZ}{\min}}&\,\, \rev{\frac{1}{T}\left( \sum\limits_{k=1}^T \|\vx_k - \mA\vz_k\|_2^2 + \lambda \|\vz_k\|_1\right) \quad\text{s.t.}\,\, \|{\va}_i\|_2 \leq 1, \forall i \in \{1, \dots, K\}}\\
   \end{aligned}
\end{equation}
where, $\|\cdot\|_{1, 1}$ denotes the sum of absolute values of the elements of the matrix, $\mX$ is the input data matrix ($\in \sR^{N\times T})$, $\mZ$ is the matrix of sparse codes, also called latent representations $(\in \sR^{K\times T})$ respectively, with $T$ being the number of independent examples, $\lambda$ is the sparsity parameter and  $\mA$ is the dictionary ($\in \sR^{N\times K})$, which maps the latent back to the input space. \rev{The $\|\cdot\|_{1, 1}$ is used to impose the sparsity constraints on latent representations $\mZ$. Such constraints are key to modeling \textit{nonlinear manifolds}, without which the model only reflects a linear subspace, spanned by the columns of the dictionary $\mA$. \Figref{fig: non-linearity} highlights this distinction, where, using sparsity constraints, one is able to model a nonlinear manifold as depicted in \Figref{fig: non-linearity}(b). Prior works have extensively used these approaches to model union of subspaces \citep{vidal2005generalized}. In neuroscience,} such models have been used to explain receptive field structure and the neuronal activity in the V1 cortex. \citep{olshausen_emergence_1996, olshausen_sparse_1997}.


While sparse dictionary learning has been used to explain the emergence of receptive fields in the V1 region of the visual cortex, predictive coding approaches provide another perspective on theory of cortical processing. Built upon the concept of sparse dictionary learning, predictive coding theory extends it to incorporate hierarchical systems in which only the essential information is propagated upwards from downstream layers, providing an efficient coding mechanism. This is achieved by prediction errors from higher (top) cortical layers of the activity in the lower(down) cortical layers, analogous to the goal of minimizing the reconstruction loss in sparse dictionary learning. Originally explored in the context of providing a theory for retinal and LGN cells by \citep{mumford1992computational}, applications of predictive coding theory to cortical functions was later extended by \citep{rao_predictive_1999}, who used it to explain sensory interactions in the visual cortex. Recent works by \citep{boutin_sparse_2021} have focused on integrating sparse dictionary learning with predictive coding to model hierarchical processing in both V1 and V2 areas of the visual cortex, while \citep{rao2024sensory} use predictive coding approaches to propose a unified theory for sensory and motor processes across the neocortex. As a result, predictive coding methods offer a promising framework for developing models that incorporate top-down interactions in the representation learning process.

In this work, we investigate how \textit{similarity based} representation learning can be adapted to a generative paradigm. 
\begin{itemize}
    \item Our first contribution is to show an equivalence between \textit{similarity} based approaches and sparse dictionary learning, \rev{extending the former to generative modeling with \textbf{overcomplete} representations. We show that learning the \textbf{latent kernel structure} in such a framework leads to an \textbf{implicit generative map} from the latent back to the input space, which are inherently \textbf{local and interpretable}, showing receptive-field like characteristics.}

    \item Our second contribution demonstrates the flexibility of our framework towards learning diverse generative maps, in particular, manifold structures. We achieve this by modifying the constraints on the learned representations and the kernel structure itself.
    \item Third, we develop a novel formulation of an Alternate Direction Method of Multipliers (ADMM) \citep{boyd_distributed_2011}) to derive the update dynamics for the variables involved in the optimization process. 
    \item Finally, we lay the foundations towards a biologically plausible framework for our proposed model, where representation learning is facilitated by combining both combining \textit{similarity based} approaches with a \textit{predictive coding} framework capturing both top-down and bottom-up interactions.
\end{itemize} 

\begin{table}[t]
\renewcommand{\arraystretch}{1.2} 
\setlength{\tabcolsep}{3pt}       
\centering
{\color{black}

\begin{tabularx}{0.85\linewidth}{|Y|Y|Y|Y|Y|}
\hline
\textbf{Approach} & \textbf{Nonlinear Manifold} & \textbf{Bioplausibility} &
\textbf{Flexible  latent dimensionality} & \textbf{Generative} \\ \hline
Principal Component Analysis (PCA)    & No  & Yes     & No  & Yes \\ \hline
Independent Component Analysis (ICA)    & No  & Yes     & No  & Yes \\ \hline
Non-negative Matrix Factorization (NMF)    & No  & Yes     & Yes & Yes \\ \hline
Non-negative Similarity Matching (NSM)    & Yes & Yes     & Yes & No  \\ \hline
\textbf{KSM-Sparsity}& \textbf{Yes} & \textbf{Yes$^*$} & \textbf{Yes} & \textbf{Yes} \\ \hline
\textbf{KSM-Manifold} & \textbf{Yes} & \textbf{No}      & \textbf{Yes} & \textbf{Yes} \\ \hline
\end{tabularx}
\vspace{1em}
\caption{Comparison of different \textit{similarity} based representation learning approaches. Here KSM-Sparsity, KSM-Manifold refers to our proposed approach with sparsity (\S \ref{sec:ksm-implicit}) and simplex constraints (\S\ref{sec: different_priors}) . $^*$Bioplausible implementations are possible when no specific constraints are imposed on the latent-space kernel structure (\S\ref{sec:bio-plausible})}
\label{tab:review-comparison-1}}

\end{table}
\
\rev{
To orient the reader, Table \ref{tab:review-comparison-1} contrasts our approach (KSM) with widely used \textit{similarity} based representation learning algorithms along four axes that are central to this work:
\begin{enumerate}
    \item Ability to model nonlinear manifolds
    \item Existence of an explicit generative map back to inputs
    \item Flexibility in latent dimensionality (over/undercomplete) and 
    \item Prospects for local/biologically plausible learning
\end{enumerate}
}
\FloatBarrier



We begin the rest of our treatment with a brief overview of classical similarity matching techniques in \S \ref{sec: similarity_matching}, highlighting their role in developing neurally plausible models for representation learning. \S \ref{sec:ksm-implicit} introduces and extends the kernel similarity matching (KSM) framework discussed previously to generative models. In \S \ref{sec: different_priors}, we explore how different priors affect representation learning under this framework, focusing on a prior inspired by metabolic constraints on neural activity. \S \ref{sec:opt} outlines the general optimization procedure for solving the KSM objective and discusses the computational considerations involved in the gradient dynamics of each optimization variable. We then present observations and insights from numerical experiments in \S \ref{sec: exp} applying the proposed frameworks to synthetic and real-world datasets. In \S \ref{sec:bio-plausible}, we take a step toward developing a neurally plausible model for representation learning. We propose a relaxed version of the KSM objective that incorporates both bottom-up and top-down interactions between neurons while enforcing local learning rules for synaptic parameters. We conclude in \S \ref{sec: conclusion} by discussing the approach's limitations and outlining future directions for contributions.


\section{Classical Kernel Similarity Matching for Representation Learning}
\label{sec: similarity_matching}

We start with a brief review of classical kernel similarity matching approaches, which have been explored in the context of developing neurally plausible architectures in recent works \citep{pehlevan_why_2018,luther_kernel_2022}. As discussed in the previous section, such a framework also underpins the extraction of meaningful representations for downstream tasks in many deep unsupervised learning architectures. 

\noindent \textbf{Problem formulation.} Mathematically, similarity based approaches to representation learning can be described as follows:
\begin{equation}
     \label{eq:similarity_matching}
   \undermin{\mZ \in \gC}\,\, d(\gK_\mX, \gK_\mZ)
\end{equation}
where, $\mX \in \sR^{N\times T}$ denotes the input data matrix (with $T$ samples each of dimension $N$) and $\mZ \in \sR^{K\times T}$ denotes the corresponding representation matrix (each of dimension $K$). $\gC$ denotes the set of constraints imposed on the representation space. Here $\gK_\mX$ and $\gK_\mZ$ are called \textit{input} and \textit{representation kernels} respectively, which capture the \textit{notion of similarity} between different samples in the dataset. More generally, for a given data or representation matrix $\mQ \in \sR^{N \times N}$ we can define $\gK_\mQ(s, t) = f_\mQ(\vq_s, \vq_t)\,\,\forall\,\,s, t \in \{1, 2, \dots T\}$ where $f_\mQ: \sR^N \times \sR^N \to \sR$ is the similarity measure between two samples $\vq_s$ and $\vq_t$ in $\sR^{N}$. A simple example for such a mapping could be $K_Q(s, t) = \vq_s^T \mathbf{Q} \vq_t$, which is a generalized inner product between $\vq_s$ and $\vq_t$, where $\mQ \succ 0$ is a positive definite matrix. If $\mQ = \mI$, we have the standard inner product. The function $d$, in \Eqref{eq:similarity_matching}, denotes a \textit{notion of distance} (need not be a strict metric) between these two kernels. Common examples for such distance measures could include the vector norms, or arccosine of the inner products. The goal of the optimization problem in \Eqref{eq:similarity_matching} being to find representations $\mZ$ which makes the input and the representation kernel similar. \citep{pehlevan_why_2018} use a standard inner product formulation of the kernel function $f$ given by $f(\vx_s, \vx_t) = \vx_s \cdot \vx_t$, which leads to the \textit{classical kernel similarity matching} problem for representation learning as given below:
\begin{equation}
    \label{eq:classical_similarity_matching}
    \begin{aligned}
    \undermin{\mZ \in \sR^{K\times T}} \frac{1}{T^2}\| \mX^T\mX - \mZ^T\mZ\|_F^2 &= \undermin{\mZ \in \sR^{K\times T}} \frac{1}{T^2}\sum\limits_{s, t}\|\vx_s^T\vx_t - \vz_s^T\vz_t\|^2\noindent\\
    &= \undermin{\mZ \in \sR^{K\times T}} \frac{1}{T^2}\left[-2\Tr(\mX^T\mX\mZ^T\mZ) + \Tr(\mZ^T\mZ\mZ^T\mZ)\right]
    \end{aligned}
\end{equation}
where the final equation comes from the fact that for any matrix $\mG$, $\|\mG\|_F^2 = \Tr(\mG^T\mG)$. 

\noindent \textbf{An online formulation to tackle computational bottleneck and improve biological feasibility.} A key bottleneck with the above formulation is that the objective scales poorly with the size of the dataset as one needs to compute the kernel matrices $\gK_\mX=\mX^T\mX$ and $\gK_\mZ = \mZ^T\mZ$ over pairs of examples for the entire dataset scaling the computation as $\gO(T^2)$. To overcome this \citep{pehlevan_why_2018} employed a Legendre transformation rewriting:
\begin{equation}
    \label{eq:legendre_transform}
    \begin{aligned}
       -\frac{1}{T^2}\Tr(\mX^T\mX\mZ^T\mZ) &\to \undermin{\mW}\, -\frac{2}{T}(\mX^T\mW^T\mZ) + \Tr(\mW^T\mW) =-\frac{2}{T}\sum\limits_{t}\vx_t^T\mW\vz_t + \Tr(\mW^T\mW)\\
       \frac{1}{T^2}\Tr(\mZ^T\mZ\mZ^T\mZ) &\to \undermax{\mM}\, \frac{2}{T}\Tr(\mZ^T\mM\mZ) - \Tr(\mM^T\mM) = \frac{1}{T}\sum\limits_{t}\vz_t^T\mM\vz_t + \Tr(\mM^T\mM)
    \end{aligned}
\end{equation}
leading to the following min-max problem:
\begin{equation}
    \label{eq:classical_similarity_matching_legendre}
    \begin{aligned}
        \undermin{\mZ \in \sR^{K\times T}} \frac{1}{T^2}\| \mX^T\mX - \mZ^T\mZ\|_F^2 = \undermin{\vz_t}\,\undermin{\mW}\,\undermax{\mM}\,\, \frac{1}{T}\sum\limits_{t}(-4\vx_t^T\mW\vz_t + 2\vz_t^T\mM\vz_t) + 2\Tr(\mW^T\mW) - \Tr(\mM^T\mM)
    \end{aligned}
\end{equation}
The key takeaway from the discussion above is that employing Legendre transformation allows for the optimization problem to be decoupled over individual samples. While we obtain an increased parameter count by introducing $\mW$ and $\mM$, the computation of the objective now scales linearly with the size of the dataset. This is also a crucial step towards biologically plausible representation learning which necessitates that neural representations are encoded in an online manner over individual samples \citep{onlinebioplausible}. 

\noindent \textbf{Extensions and limitations.} In \citep{luther_kernel_2022}, the authors extend this framework to a more general kernel structure for $f$ in the input space ($\mX$) while using the standard inner product kernel for the latent space ($\mZ$) \rev{and in \citep{sengupta_manifold-tiling_2018}, the authors impose a non-negativity constraint on the latent to learn locally selective representations along the input manifold}. However, these formulations of the kernel matching objective, typically encode representations in a bottom-up manner (from lower to higher levels) only, neglecting any feedback from higher to lower levels. As a result, they lack top-down interactions that map the representations back to the input space, a key aspect suggested by prevalent theories of cortical processing which view perception as a \textit{controlled hallucination} where the implicit model that the brain uses is corrected by feedback of the sensory prediction error \citep{deneve2016circular, millidge2021predictive}. In the next section we articulate how these similarity based approaches can be extended to learn representations from the input as well as learning an implicit generative map that links them back to the input space.


\section{Kernel Similarity Matching (KSM) as implicit generative models}
\label{sec:ksm-implicit}
To extend the kernel matching framework for generative models, we focus on dictionary learning methods, where representations are learned by minimizing the prediction error with respect to the input while having some form of constraints on the latent space and show an equivalence between the two.

\noindent \textbf{A similarity-matching formulation of sparse dictionary learning.} We begin with a modified version of the sparse dictionary learning objective (described in Equation \ref{eq:orig_sparse_coding}), which incorporates the norm constraints on the dictionary columns directly into the objective itself as demonstrated in \citep{lee2006efficient, tolooshams_stable_2022}. From an optimization standpoint, the norm constraints on the dictionary columns solve the scaling ambiguity in the objective in \Eqref{eq:orig_sparse_coding} where the sparse representations and the dictionary columns can be scaled arbitrarily while maintaining the same reconstruction loss. This improves the stability and convergence of the optimization approaches on a variety of sparse dictionary learning based models \citep{jiang2021improved}. From a neuroscience perspective, normalized dictionary columns ensure that the learned codes have a consistent scale of contribution towards encoding the input, with the magnitude of contribution directly influenced by the amplitude of the corresponding code. This facilitates better interpretability when analyzing population activity across different regions of the brain \citep{tolooshams2024interpretable}. Furthermore, normalization of dictionary columns aligns with the concept of synaptic normalization \citep{turrigiano2004homeostatic,mayzel2023homeostatic} in neuroscience, where an increase (decrease) of a particular synaptic weight, increases (decreases) the strength of other synapses in the network, leading to a more biologically realistic theory of networks encoding neural population codes. The modified sparse coding objective given below incorporates the norm constraints on the dictionary columns as a penalty or regularizer in the objective itself which leads to the revised optimization problem:
\begin{equation}
    \label{eq:sc_dea}
    P_1: \underset{\mA, \mZ}{\min}\,\,\frac{1}{2T}\|\mX - \mA\mZ\|^2_F + \frac{\lambda}{T}\|\mZ\|_{1, 1} + \frac{\omega}{2} \|\mA\|_F^2 \qquad \omega > 0
\end{equation}
Here, $\|\mZ\|_{1, 1} = \sum\limits_{i, j}|z_{ij}|$. 


The objective in Equation \ref{eq:sc_dea} is a \textit{bilevel optimization} problem which is traditionally solved in literature \citep{olshausen1996emergence}, \citep{dempe2020bilevel}, \citep{zhang2024introduction} through alternating minimization, where one solves the objective first w.r.t. to the latent representation $\mZ$ (inner optimization problem) while fixing $\mA$ to obtain the optimum $\mZ^{*}$. Thereafter, fixing $\mZ$ to $\mZ^*$, we solve the objective w.r.t. $\mA$ (outer optimization problem), and repeat these two steps until convergence. To see the connection to kernel similarity matching, we make use of the Lemma below on the bilevel objective function. We include the proof of the lemma in the appendix for completeness.
\begin{lemma}[Min-Min Lemma]
    \label{lem: Min-Min Lemma}
    Let $f(\vx, \vy)$ be a function with finite minima where $\vx \in \sR^m$ and $\vy \in \sR^n$ for $m, n \in \sN$ respectively, then the following holds:
    \begin{align*}
        \undermin{\vx}\,\,\undermin{\vy}\,\,f(\vx, \vy) = \undermin{\vy}\,\,\undermin{\vx}\,\,f(\vx, \vy)
    \end{align*} 
\end{lemma}
\begin{proof}
    See Appendix \ref{app: min-min-lemma}
\end{proof}
Essentially, Lemma \ref{lem: Min-Min Lemma} allows us to switch the order of optimization in a bilevel optimization problem. Consequently, our inner optimization problem in \Eqref{eq:sc_dea} now becomes w.r.t. $\mA$ followed by $\mZ$ for the outer optimization. \rev{Solving the inner optimization w.r.t $\mA^*$ to get $\mA^* = \frac{\mX\mZ^T}{T}\left(\frac{\mZ\mZ^T}{T} + \omega \mI\right)^{-1}$ and plugging it back in the objective for the outer optimization in $\mZ$ leads us to the following proposition}:
\begin{proposition}
    \label{prop: ksm-obj}
    Let $P_1$ be the sparse dictionary learning problem as described above and $Q_1$ be the formulation as described below:
    \begin{equation}
        \label{eq:ksm-sc}
        \begin{aligned}
            Q_1: \underset{\mZ, \mH}{\min}\,\,& -\frac{1}{2T^2}\Tr(\underbrace{\mX^T\mX}_{\gK_{\mX}}\underbrace{\mZ^T\mH^{-1}\mZ}_{\gK_{\mZ}}) + \frac{\lambda}{T} \|\mZ\|_{1, 1}\\
          \text{s.t.}\,\,& \mH = \frac{\mZ\mZ^T}{T} + \omega\mI\\
        \end{aligned}
    \end{equation}
    then from lemma \ref{lem: Min-Min Lemma} we have $P_1$ and $Q_1$ are equivalent.
\end{proposition}

\begin{proof}
    See Appendix \ref{app: prop-ksm-obj}
\end{proof}

\textbf{Deconstructing the objective.} The objective in \Eqref{eq:ksm-sc} consists of a trace term and a $L_1$ norm regularizer for imposing sparsity on the representations. Additionally, a correlation constraint is enforced on the representations by the matrix $\mH$. The trace term represents the matrix inner product between a standard linear input kernel $\gK_\mX = \mX^T\mX$ and a modified representation kernel $\gK_\mZ = \mZ^T\mH^{-1}\mZ$, where the structure of the representation kernel is governed by the matrix $\mH$, rather $\mH^{-1}$. Notably, if $\mH = \mI$ then we are back to the standard inner product kernel in the representation space (given by $\mZ^T\mZ$). 

To gain some intuition about the latent kernel function $\gK_\mZ$ and the corresponding structure matrix $\mH^{-1}$, we observe that the matrix $\mH$ as defined by the constraint in \Eqref{eq:ksm-sc} mimics the sample covariance of the learned representations. Therefore, the transformation $\mH^{-\frac{1}{2}}\mZ$, essentially de-correlates the latent representation $\mZ$, and $\mK_\mZ$ computes the kernel value in this decorrelated latent space. In essence, the trace term, which denotes the matrix inner product between $\gK_\mX$ and $\gK_\mZ$,  captures the similarity measure between the input kernel and the representation kernel. Minimizing the objective in \Eqref{eq:ksm-sc} is equivalent to maximizing the trace inner product subject to the sparsity and correlation constraints defined in the problem. This, in turn, implies making the input and the representation kernels similar. Juxtaposing Equation \ref{eq:ksm-sc} with the \textit{similarity} based representation learning described in Equation \ref{eq:similarity_matching}, we observe that the sparse dictionary learning objective (\Eqref{eq:sc_dea}) is equivalent to kernel matching representation learning problem, with an imposed structural constraint on the representation kernel. 

This formulation of kernel matching differs from prior similarity-based approaches in one key aspect. In prior methods \citep{pehlevan_why_2018,luther_kernel_2022}, the representation kernel involved in the similarity matching objective was typically restricted to a standard inner product-based formulation. In contrast, the representation kernel in this case is derived from a generalized inner product formulation with the structure of the inner product governed by the matrix $\mH^{-1}$, which itself is being estimated through a constraint in the optimization problem, allowing the entire kernel structure to evolve as the representations $\mZ$ are being learned. We argue that it is this flexibility in the kernel structure that allows us to formulate an implicit generative map, tying the representations back to the input space.

\textbf{An implicit generative map.} 
As discussed earlier, the kernel matching objective presented in \Eqref{eq:ksm-sc} learns representations $\mZ$ corresponding to the input samples $\mX$ by aligning the input and the representation kernel. This also allows us to calculate an \textit{implicit generative map} from the representation space to the input space by estimating the dictionary $\hat{\mA}$, where, $\hat{\mA} =  \dfrac{\mX\mZ^T\mH^{-1}}{T}$ (See Appendix \ref{app: prop-ksm-obj} for the proof of this relation). The model prediction $\hat{\vx}$ for the corresponding input $\vx$ can then be calculated as $\hat{\vx} = \hat{\mA}\vz$, where $\vz$ is the representation of $\vx$ in the latent space with the final expression for $\hat{\vx}$ being given as $\hat{\vx} = \dfrac{\mX\mZ^T\mH^{-1}}{T}\vz = \dfrac{1}{T}\sum\limits_{i=1}^T \vx_i \langle \vz_i, \vz \rangle_{\mH^{-1}}$, where $\langle \cdot, \cdot \rangle_{\mH^{-1}}$ computes the representation kernel between $\vz_i$ and $\vz$, with the structure of the kernel being determined by $\mH^{-1}$. 



\section{A flexible framework with different priors}
\label{sec: different_priors}
The representation learning problem articulated in \Eqref{eq:similarity_matching} provides a flexible framework that varies depending on the prior imposed on the latent space. Specifically, in the previous section we discussed how imposing the sparsity constraint (using the $L_1$ norm) on the latent space alongside the correlation constraint on the representation kernel $\mH$ is equivalent to a sparse dictionary learning problem. Additionally, other sparsity based priors \citep{bach2012optimization} on the latent representations that do not involve explicit equality constraints also amount to the same formulation as discussed before. It is worth noting that the correlation constraint on the representation kernel $\mH$ is a consequence of introducing a norm constraint (an $L_2$ prior) on the columns of the dictionary ($\mA$)  and the reconstruction loss, as mentioned in \Eqref{eq:sc_dea}, which in turn leads to Proposition \ref{prop: ksm-obj}. In this section, we explore how the optimization problem in \Eqref{eq:ksm-sc} can be extended to other forms of $L_2$ like priors on the dictionary columns which can lead to different behaviors.

\textbf{A probability simplex prior.} 
We focus our discussion on an interesting case where the representations are constrained on a probability simplex, i.e. each dimension is non-negative and sum across all the dimensions are constrained to unity. From a probabilistic perspective, this corresponds to imposing a \textit{dirichlet prior} on the representations.

\textbf{Why this prior?} Our choice of this prior is motivated by metabolic constraints in a biological system. In neuroscience, imposing the sparsity constraint is desirable for representation learning as it is tied to minimizing wiring length \citep{foldiak2003sparse} and metabolic efficiency for neural activity \citep{baddeley1996efficient}, with the latter arguing the fixing the mean firing rate leads to sparsity in the representations. While the $L_1$ norm constraint imposes sparsity, it does not strictly limit the amplitude of each of the desired representations. Imposing sparsity while at the same time limiting the amplitude of the representations can be a more desirable way of imposing metabolic constraints from a biological perspective, as it imposes competition in the population by forcing activity of each neuron (which corresponds to a latent dimension) is bounded, and is undergoes subtractive normalization by the total population activity (i.e. each of the dimensions add upto unity). The idea of normalization of individual neuronal responses by the population response has been formulated as a canonical form of neural computation in different sensory inputs \citep{carandini2012normalization, chalk2017sensory} across different species. \citep{carandini2012normalization} suggest that this kind of computation plays a role in making neural responses more invariant to stimulus properties like contrast or intensity, while at the same time reducing redundancy in the learned representations, whereas \citep{chalk2017sensory} argue that the type of normalization is an inherent feature of the kind of noise associated with the generated response to the input stimulus. For our case, the simplex prior, introduces a subtractive normalization effect on neuronal responses where the inhibition in individual activity is additive in nature and governed by the sum of the population activity (See Appendix \ref{app: proj-simplex} for details).

To investigate such a prior on the latent, we start with a dictionary learning objective proposed in a prior work by \citep{tasissa_k-deep_2023}, where the alongside the representations being constrained on a probability simplex, the proximity of the dictionary columns to the data points is also emphasized in addition to the reconstruction loss. The optimization objective for this case can then be articulated as follows:

\begin{equation}
    \label{eq:manifold_learning}
    \begin{aligned}
       P_2: \underset{\mA, \mZ}{\min}&\,\, \frac{1}{2T}\|\mX - \mA\mZ\|_F^2 + \frac{\omega}{T}\sum\limits_{k=1}^K\sum\limits_{l=1}^T z_{kl}\|\vx_l - \va_k\|_2^2\\
       \text{s.t.}&\sum\limits_{k=1}^K z_{kl} = 1, \,\, z_{kl} \ge 0, \quad \forall \quad l\,\in\,\{1, 2, \dots, T\}
    \end{aligned}
 \end{equation}
 \textbf{Deconstructing the regularizer. }The regularizer here plays an important role in ensuring only dictionary columns $\va_j$ that are close to the input sample $\vx_i$ are utilized in minimizing the reconstruction error (as when $z_{kl}$ is high $\va_k$ needs to be close to $\vx_l$ to minimize the penalty). Additionally, as the representations lie on a probability simplex, the predicted samples by the model lie within the convex hull of the dictionary columns. As a result the objective in \Eqref{eq:manifold_learning} is able to piece-wise tile the manifold of the input data with convex polygons, with the dictionary columns $\va_j$ serving as \textit{anchor points} or vertices of these convex polygons \citep{tasissa_k-deep_2023}. Consequently, these dictionary columns serve as localized receptive fields tiling the surface of the input manifold, with neuronal activity tuned to the location of the receptive fields along the manifold surface. This can be helpful in understanding how a neural manifold can be decoded by downstream system for different task characteristics \citep{jazayeri2021interpreting, chung2021neural}. For instance, spatially localized response is a key mechanism by which hippocampus place cell facilitate navigation \citep{o1978hippocampus}. In addition, head direction circuits in fly also involve localized receptive fields that are oriented to the angle where the head is pointing towards \citep{chaudhuri2019intrinsic}. Other works have also explored the role localized receptive fields play in understanding input features such as object and orientation selectivity in the visual cortex \citep{hubel1959receptive, chung2021neural} and frequency selectivity in the auditory cortex \citep{knudsen1978center,da2013tuning}.

 In later sections, we show that imposing such simplex based constraints as shown in \Eqref{eq:manifold_learning_kernel} not only allows us to learn localized receptive fields along the surface of the manifold, but together with the estimated latent representations, piece-wise approximate the input data manifold, with the localized receptive fields serving as \textit{anchors} on the input manifold, and the latent representations modeling population activity tuned to these receptive fields over different sections of the input manifold.

 \textbf{Kernel Similarity Matching (KSM) for Manifold Learning.}
 We note that the objective in \Eqref{eq:manifold_learning} is again a bilevel optimization problem in $\mA$ and $\mZ$. Therefore, we once again employ lemma \ref{lem: Min-Min Lemma} to switch the order of optimization, \rev{using $\mA^* = (1+\omega)\left(\frac{\mZ\mZ^T}{T} + \omega \mD\right)$, with $\mD = \text{diag}\left(\frac{\mZ \bm{1}_T}{T}\right)$ for the outer objective, }which leads to the following proposition:

 \begin{proposition}
    \label{prop: ksm-manifold}
    Let $P_2$ be the optimization problem as described before and $Q_2$ be the formulation as described below:
    \begin{equation}
        \label{eq:manifold_learning_kernel}
        \begin{aligned}
           Q_2: \underset{\mZ, \mH}{\min}\,\,& -\frac{1}{2T^2}\Tr(\mX^T\mX\mZ^T\mH^{-1}\mZ)\\
             \text{s.t.}\,\,&\,\, \mZ^T\mathbf{1}_K = \mathbf{1}_T\\
             &\,\,\mZ \ge 0\\
             &\,\,\mH = \frac{\mZ\mZ^T}{T} + \omega\mD  \,\, \text{where}\,\,\mD = \text{diag}\left( \frac{\mZ \mathbf{1}_T}{T} \right)
        \end{aligned}
     \end{equation}
     Then $P_2$ and $Q_2$ are equivalent.
\end{proposition}

\begin{proof}
    See Appendix \ref{app: prop-ksm-manifold}
\end{proof}
We see that the objective in \Eqref{eq:manifold_learning_kernel} aligns with the structure of kernel similarity matching described in \Eqref{eq:similarity_matching}, where representations are learned by maximizing the notion of similarity as captured by trace inner product between the standard input kernel (described by $\mX^T\mX$) and a generalized structured representation kernel (described by $\mZ^T\mH^{-1}\mZ$), with the representations $\mZ$ being constrained to the probability simplex and the kernel structure determined by the matrix $\mH$ , where $\mH = \dfrac{\mZ\mZ^T}{T} + \omega \mD$, $\mD$ being a diagonal matrix, indicating the ``degree'' of involvement of the corresponding dictionary column across different samples in the dataset. Similar to the sparse dictionary case in the previous section, the objective in \Eqref{eq:manifold_learning_kernel} learns representation while learning an implicit generative map given by $\hat{\mA^*} = (1+\omega)\dfrac{\mX\mZ^T\mH^{-1}}{T} $ (see Appendix \ref{app: prop-ksm-manifold} for details) which depends on the structure of the representation kernel $\mH$ and the latent representations $\mZ$. Similar to the previous section, the model prediction $\hat\vx$ for the corresponding input $\vx$ can be given as $\hat\vx = \hat\mA\vz$, where $\vz \in \Delta^K$ is the representation of $\vx$ in the latent space, and $\Delta^K$ denotes the probability simplex in $K$ dimensions. Thus, we can write $\hat\vx = \hat\mA\vz = \dfrac{(1+\omega)}{T} \sum\limits_{i=1}^T \vx_i \langle \vz_i, \vz\rangle_{\mH^{-1}}$, where $\langle \cdot, \cdot \rangle_{\mH^{-1}}$ computes the representation kernel between $\vz_i$ and $\vz$, with the structure of the kernel being determined by $\mH^{-1}$. Unlike prior works on locality preserving feature learning \cite{luther_kernel_2022}, the approach presented here in addition to learning local representations (due to sparsity and competition, see \S\ref{sec: exp-synthetic-simplex}) implicitly learns a generative map from the learned representations. We demonstrate this approach on a simulated dataset in Section \ref{sec: exp} thereby showing the flexibility of the kernel matching framework in \Eqref{eq:similarity_matching} to learn diverse generative maps by modifying the constraints on the latent space and the representation kernel structure itself. 



\section{Optimization Procedure}
\label{sec:opt}
In this section, we discuss the basic approach to solving the optimization problems described in Equations \ref{eq:ksm-sc} and \ref{eq:manifold_learning_kernel}. While online optimization approaches have been previously developed for kernel matching objectives \citep{pehlevan_why_2018, luther_kernel_2022}, little has been explored about solving a kernel matching objective in which the representation kernel is adaptive, with constraints on both the latent representation and the latent kernel structure itself. A popular method for solving any constrained optimization problems, such as those described by Equations \ref{eq:ksm-sc} and \ref{eq:manifold_learning_kernel}, is the Alternating Direction Method of Multipliers (ADMM) \citep{boyd_distributed_2011}. We provide a brief primer for the ADMM algorithm below and show how it can be adapted to solve the objective in our case.

\textbf{The ADMM algorithm.} ADMM is a class of optimization algorithms that are particularly well suited for solving optimization problems where the objective is composed of separable entities each comprising different optimization variable. The algorithm builds upon the dual ascent algorithm \citep{boyd_distributed_2011} which involves first optimizing over the principle optimization variables (otherwise known as \textit{primal variables}) in the \textit{augmented Lagrangian} (Lagrangian with a $L_2$ penalty imposed on the constraint term) to obtain the dual objective followed by gradient ascent on the dual variable (also known as the \textit{Lagrange multiplier}).  The ADMM algorithm takes a step further by performing sequential optimization of each of the \textit{primal variables}. This in turn breaks the optimization problem into smaller sub-problems over individual optimization variables which can be solved more easily and independently. This approach becomes highly effective if the objective and the constraint itself are separable (can be decomposed into sum of terms each involving a different optimization variable) as each of these sub-problems can be solved in parallel. While convergence guarantees typically solicit convexity assumption on the objective, ADMM methods, in practice, tend to work well even for non-convex objectives \citep{xu2016empirical,hajinezhad2016nonnegative,wang2019global}, thereby making them a popular choice for solving constrained optimization problems.

\textbf{An ADMM formulation for KSM.} As the kernel matching formulation discussed before (\Eqref{eq:sc_dea}) is in principle a constrained optimization problem, we can put the words in the previous paragraph to practice and formulate an ADMM based approach to solve the optimization problem in Equations \ref{eq:ksm-sc} and \ref{eq: ksm-manifold-scaled}. The \textit{primal variables} of optimization are the latent representations $\mZ$. The representation kernel matrix $\mH$ and the constraints are incorporated in the objective with the help of the \textit{dual variable} (Lagrange multiplier) $\mM$. Algorithm \ref{algo:admm} describes how the sequential optimization over $\mZ$ and $\mH$ would proceed alongside the update for the Lagrange multiplier $\mM$. While we focus on the sparse representation learning in the upcoming discussions, the same can be extended to the problem of manifold learning (Appendix \ref{app: subsec-manifold}).


\begin{algorithm}
    \caption{An ADMM approach to model the optimization dynamics}
    \KwIn{data matrix $\mX$}
    \KwOut{latent matrix $\mZ$ , latent kernel matrix $\mH$}
    \For{$k = 1, 2, \dots$}
    {
        $\mZ^{k+1} \gets \underset{\mZ}{\arg\min}\,\,\Ls(\mZ, \mH^{k},\mM^k)$\\
        $\mH^{k+1} \gets \underset{\mH}{\arg\min}\,\,\Ls(\mZ^{k+1}, \mH, \mM^k)$\\
        $\mM^{k+1} \gets \mM^k + \rho\left(\frac{\mZ^{k+1}\mZ^{k+1^T}}{T} + \omega \mI - \mH^{k+1}\right)$

    }
    \label{algo:admm}
\end{algorithm}
Here $\Ls(\mZ, \mH, \mM)$ is the augmented Lagrangian given by:
\begin{equation}
    \label{eq:aug_lag}
    \begin{aligned}
    \Ls(\mZ, \mH, \mM) &= -\frac{1}{2T^2}\Tr(\mX^T\mX\mZ^T\mH^{-1}\mZ) + \frac{\lambda}{T} \|\mZ\|_{1, 1} \\
    &+ \Tr\left(\mM^T\left(\frac{\mZ\mZ^T}{T} + \omega\mI - \mH\right)\right) + \frac{\rho}{2}\left\|\frac{\mZ\mZ^T}{T} + \omega\mI - \mH\right\|_F^2
    \end{aligned}
\end{equation}
with Algorithm \ref{algo:admm} solving the min-max problem $\underset{\mZ, \mH} {\min}\,\underset{\mM}{\max}\,\,\Ls(\mZ, \mH, \mM)$. However, the objective defined in \Eqref{eq:aug_lag} does not scale well with the size and dimensionality as the computation of the kernel in the objective scales as $NT^2$, where $N$ denotes the data dimensionality and $T$ denotes the number of samples in the dataset. This is also an obstacle to development of biologically plausible algorithms, since it requires access to all the samples in the dataset to compute the objective function and learn the corresponding representations, which deviates from the idea that our brain encodes representations in an online manner as discussed earlier in Section \ref{sec: similarity_matching}.

\textbf{Towards a separable formulation.}
 To overcome the challenge described before, we employ Legendre transforms discussed in Section \ref{sec: similarity_matching} on the trace term in \Eqref{eq:ksm-sc} to obtain a separable formulation across different sample points of the objective in \Eqref{eq:ksm-sc}. The resultant objective is given as:

\begin{equation}
    \label{eq:sc_dea_kernel_sep}
    \begin{aligned}
    \undermin{\mW, \mZ, \mH}&\,\,-\frac{1}{T}{\Tr\left(\mX^T \mW^T \mZ\right)} + \frac{1}{2}\Tr\left( \mW^T\mH\mW \right) + \frac{\lambda}{T}\|\mZ\|_{1,1}\\
    \text{s.t.}&\,\,\mH = \frac{\mZ\mZ^T}{T} + \omega \mathbf{I}
    \end{aligned}
\end{equation} 
where, we break down the trace similarity measure between the input and the representation kernels in \Eqref{eq:ksm-sc} into two separate trace terms using an auxiliary variable $\mW$ as shown below:
\begin{equation}
    \label{eq:legendre_transform_sep}
    \dfrac{1}{2T}\Tr(\mX^T\mX\mZ\mH^{-1}\mZ) = \undermin{\mW} -\frac{1}{T}{\Tr\left(\mX^T \mW^T \mZ\right)} + \frac{1}{2}\Tr\left( \mW^T\mH\mW \right)    
\end{equation}

The first trace term on the right-hand side of the above equation now can be separated across individual samples, $\vx_i$, and their corresponding representations,  $\vz_i$. However, we note that the incorporating the constraint to formulate the Lagrangian would still require access to all the samples from the given dataset, which would imply that the augmented Lagrangian described by \Eqref{eq:aug_lag} cannot be separated over samples. Our strategy here is to split the constraint over different samples. Essentially, we break down the kernel matrix $\mH$ as a mean over $T$ separate matrices $\mP_i$ each of which is constrained to $\vz_i\vz_i^T + \omega \mI$ such that $\mH = \frac{1}{T}\sum\limits_{i=1}^T \mP_i$. Consequently, our objective now has $T$ constraints instead of one, and rewriting Equation \ref{eq:sc_dea_kernel_sep} with the new constraints leads to the following proposition:

\begin{proposition}
    \label{prop: online-obj-ksm-sc}
    Let $R_1$ be the optimization problem as described below. Then $R_1$ is equivalent to $Q_1$.
    \begin{equation}
        \label{eq:form1c}
        \begin{aligned}
        R_1: \undermin{\mW, \vz_i, \mP_i}&\,\,-\frac{1}{T}\sum\limits_{i=1}^T \left(\vx_i^T \mW^T \vz_i\right) + \frac{1}{2T}\sum\limits_{i=1}^T \Tr\left( \mW^T\mP_i\mW \right) + \frac{\lambda}{T}\sum\limits_{i=1}^T\|\vz_i\|_{1}\\
        \text{s.t.}&\,\,\mP_i = \vz_i\vz_i^T+ \omega \mathbf{I} \qquad \forall\,\,i \in \{1, 2, \dots, T\}
        \end{aligned}
    \end{equation}
\end{proposition}

\textbf{Splitting the constraints leads to a fully separable objective.} This formulation in Equation \ref{eq:form1c} allows us to write the augmented Lagrangian in \Eqref{eq:aug_lag} in a fully separable form. This in turn helps to deal with large dataset as we can use online or parallel optimization for optimizing sample dependent optimization variables and use batch approximations for sample independent optimization terms. The revised augmented Lagrangian for the separable optimization problem in \Eqref{eq:form1c} is given by:
\begin{equation}
    \label{eq:aug_lag_sep}
    \begin{aligned}
    \bm{\gL} \left(\mZ, \mP, \mW, \mM\right) &= \frac{1}{T}\sum\limits_{i=1}^T \Ls_i \left(\vz_i, \mP_i, \mW, \mM_i\right)\\
    \text{where}, \quad\Ls_i &= -\vx_i^T\mW\vz_i + \frac{1}{2}\Tr(\mW^T\mP_i\mW) + \lambda\|\vz_i\|_1 + \rho\left\|\vz_i\vz_i^T + \omega \mI - \mP_i + \frac{\mM_i}{\rho}\right\|_F^2    
    \end{aligned}
\end{equation}
where, $\vz_i, \mP_i, \mW$ serve as the \textit{primal variables} of the optimization problem and $\mM_i$ is the \textit{dual variable} or the \textit{Lagrange multiplier}.


\subsection{Extending the ADMM algorithm to solve the online objective}
We now update the ADMM optimization steps in Algorithm \ref{algo:admm} to optimize the augmented Lagrangian in \Eqref{eq:aug_lag_sep}. The revised steps for the optimization process are outlined in Algorithm \ref{algo:admm_sep}.

\begin{algorithm}
    \caption{A novel ADMM procedure for solving Equation \ref{eq:ksm-sc}}
    \KwIn{data matrix $\mX = [\vx_1, \vx_2, \dots \vx_T], \vx_i \in \sR^{N}$}
    \KwOut{latent matrix $\mZ = [\vz_1, \vz_2, \dots \vz_T]$ , $\mP_i\,\,\forall\, i \in \{1,2,\dots, T\}$}
    \For{$k = 1, 2, \dots$}
    {
        \For{$i = 1, 2, \dots T$}
            {
                $\vz_i^{k+1} \gets \underset{\vz}{\arg\min}\,\,\Ls_i(\vz_i, \mP_i^{k},\mM_i^k)$\quad (Z-Step)\\
                $\mP_i^{k+1} \gets \underset{\mP}{\arg\min}\,\,\Ls_i(\vz_i^{k+1}, \mP, \mM_i^k)$\quad (P-Step)\\ 
                $\mM_i^{k+1} \gets \mM_i^k + \rho\left(\dfrac{\vz_i^{k+1}\vz_i^{k+1^T}}{T} + \omega \mI - \mP_i^{k+1}\right)$ \quad (M-Step)
            }
            $\mW^{k+1} \gets \underset{\mW}{\arg\min}\,\,\frac{1}{T}\sum\limits_{i=1}^{T}\Ls_i(\vz_i^{k+1}, \mP_i^{k+1}, \mW, \mM_i^k)$ \quad (W-Step)\\
    }
    \label{algo:admm_sep}
\end{algorithm}

\subsubsection{Unpacking the algorithm} 
The optimization process in Algorithm \ref{algo:admm_sep} consists of four steps that correspond to optimizing the augmented Lagrangian w.r.t. each of the primal variables $\vz_i$ (\textit{Z-Step}), $\mP_i$ (\textit{P-Step}), $\mW$ (\textit{W-Step}) and dual variables, $\mM_i$ (\textit{M-Step}), respectively.

\textbf{Separability allows parallel optimization.} The separable formulation of the augmented Lagrangian in \Eqref{eq:aug_lag_sep} allows us to perform the \textit{Z-Step, P-Step, M-Step} in parallel. However, the primal variable $\mW$ is shared across all the samples and consequently, the \textit{W-Step} requires computing the augmented Lagrangian over the entire step. We workaround this challenge by performing the \textit{W-Step} (primal variable updated) after the \textit{M-Step} (dual variable update) in the optimization process which differs from standard ADMM approaches, where, the \textit{dual variable} ($\mM_i$) updates typically follows the updates of the \textit{primal variables} of the optimization. We are able to do this because, in this case the update dynamics for the \textit{M-Step} are independent of the \textit{W-Step} and vice-versa (Appendix \ref{app: subsec-ksm-sc-dynamics}), allowing us to perform the \textit{W-Step} after the \textit{M-Step} without affecting the optimization process. This allows us to perform the sample dependent steps namely the \textit{Z-Step}, \textit{P-Step} and the \textit{M-Step} in parallel across different samples before updating the sample independent variable in the \textit{W-Step}, where we use mini-batch approximations to accelerate the optimization process.

Of the various steps outlined in Algorithm \ref{algo:admm_sep}, the \textit{P-Step} has a tractable closed form solution, and along with the \textit{M-Step} is computationally efficient. However, Both the \textit{Z-Step} and the \textit{W-Step} involve non-tractable or computationally inefficient closed form solutions. In the next section, we discuss some computational considerations for these two steps next that allow us to tackle this obstacle.

\label{sec: computational_considerations}
\subsubsection{Gradient Based updates for \textit{Z-step} and \textit{W-Step}:} 
The \textit{W-Step} as described in Algorithm \ref{algo:admm_sep} is the only sample independent step (i.e. the variable $\mW$ is the same across all the samples) and hence, requires the entire dataset to compute the objective. We choose simple gradient descent updates for the \textit{W-Step} as it prevents the need for matrix inversion in a closed form solution while allowing us to approximate the gradient computation over the entire dataset with a small batch of samples, thereby reducing the computation time. 

The \textit{Z-step} in the algorithm is of particular interest as it involves optimizing a \textit{quartic} function in the variable $\vz_i$.  This makes finding a tractable closed-form expression for the optimization sub-problem in the \textit{Z-step} non-trivial. Our approach relies on proximal gradient methods to iteratively optimize the Lagrangian, which involves a non-smooth function in the form of the $L_1$ norm, w.r.t $\vz$ for the \textit{Z-Step}. A convenient alternative for optimizing the non-smooth function is the use sub-gradients. Sub-gradients, though are easy to compute for the $L_1$ norm, are inherently more noisy due to the stochastic nature of the sub-gradient of the non-smooth functions \citep{tibshiranicnvx}.

Proximal gradient methods are quite popular for optimizing functions that are involve a smooth differentiable component (e.g. $h(\vz) = \frac{1}{2}\|\vx - \mA\vz\|_2^2$) and a non-smooth component (e.g. $g(\vz) = \lambda\|\vz\|_1$), such as the one described in Equation \ref{eq:orig_sparse_coding} \citep{bredies2008linear}. These approaches rely on the gradient of the smooth function being \textit{Lipschitz continuous}, with the Lipschitz constant given by $L = \sigma_{\max}(\mA^T\mA)$ i.e. the maximum singular value of $\mA^T\mA$ for the case described before. $L$, being independent of $\vz$, acts as a uniform upper bound for the \textit{Hessian} of the smooth function $h(\vz)$. Furthermore, $L$ directly determines the \textit{step size}, $\eta_Z$, for the gradient updates in proximal gradient methods, with $\eta_Z = \frac{1}{L}$. 

Proximal gradient methods work well when the smooth function in the objective tends to be \textit{quadratic} in the function variable, allowing $L$ to uniformly bound the Hessian. However, in the \textit{Z-Step}, the smooth component of the Lagrangian $\gL_i$, is a \textit{quartic} function in $\vz_i$, which makes the calculation of a Lipschitz constant to uniformly bound the gradient non-trivial, making it difficult to deploy proximal gradient methods. Consequently, we articulate below different optimization schedules, which involves changing the order of how \textit{Z-Step}, \textit{P-Step} and \textit{M-Step} computations are performed, allowing us to deploy proximal gradient methods for accelerating the optimization of the Lagrangian in the \textit{Z-Step}.

\subsection*{Implementing proximal gradient descent for quartic function through different optimization schedules:}

We propose three different schedules for the optimization process outlined in Algorithm \ref{algo:admm_sep} changing the order of how $\textit{Z-Step}$, $\textit{P-Step}$ and $\textit{M-Step}$ are performed. The schedules are as follows:

\begin{itemize}
        \item \underline{\textit{Schedule 1}}: In this schedule, we perform \textit{P-step} after every gradient step of the \textit{Z-step}. Essentially, the variable $\mP_i$ tracks the updates to $\vz_i$. In addition, we perform a modified \textit{M-Step} after the \textit{W-Step} by setting $\mM_i^{k+1} \gets \frac{\mW^{k+1}\mW^{k+1\,T}}{2}$.

        \item \underline{\textit{Schedule 2}}: In this schedule, we perform the \textit{P-Step} after every gradient step of the \textit{Z-Step} and the updates for the rest follows Algorithm \ref{algo:admm_sep}.

        \item \underline{\textit{Schedule 3}}: In this schedule, we perform the optimization as described in Algorithm \ref{algo:admm_sep}.

\end{itemize}

\subsubsection{Equivalence of different Schedules to Algorithms solving Sparse Dictionary Learning (SDL):}
To shed some light into the optimization dynamics undertaken by the different schedules of optimization, we propose the following theorem that draws a link between Algorithm \ref{algo:admm_sep} and conventional Sparse Dictionary Learning Algorithms involving Iterative Soft Thresholding Algorithm (ISTA) \citep{daubechies2004iterative} sparse coding and alternate minimization for Dictionary Learning \citep{lee2006efficient}.

\begin{theorem}[\underline{\textbf{ISTA based alternate minimization for SDL (Schedule 1)}}]
     When \textit{P-step} is performed after every gradient step of the \textit{Z-step} in \textit{Schedule 1}, the optimization dynamics of the objective in \Eqref{eq:sc_dea_kernel_sep} follows the Iterative Soft Thresholding Algorithm (ISTA) for sparse coding\citep{daubechies2004iterative}. Furthermore, the updates to $\mW$ together with $\vz_i$ in \textit{Schedule 1} performs alternate minimization for sparse dictionary learning. 
\end{theorem}

\begin{proof}
    See Appendix \ref{app: subsubsec-equi-ISTA}
\end{proof}

\underline{\textit{Delayed ISTA for SDL (Schedule 2)}}: Under \textit{Schedule 2}, the argument of the soft-thresholding function (\Eqref{eq: z-step}, Appendix \ref{app: prop-ksm-obj}) is still linear $\vz_i$ and closely mimic ISTA as before, though there are some deviations early on due to the M-Step updates, which subside as the \textit{M-steps} reach a fixed point. Consequently, we chose the same value for $\eta_Z$ as \textit{Schedule 1}. 

\underline{\textit{Generalized Alternate Minimization for SDL (Schedule 3)}}: Finally, under \textit{Schedule 3}, the argument of the soft-thresholding function is no longer linear and as a result, the Hessian of the smooth function is no longer uniformly bounded. As a result, we need larger values of the Lipschitz constant $L$ to upper bound the Hessian, creating a much more relaxed upper bound of the objective which leads to $\eta_Z$ being set to a significantly smaller value ($\sim 1/100)$ compared to the previous schedules for the proximal gradient steps to work. To avoid tuning  $\omega$ in all the cases described before, we normalized $\mW$ to have unit norm rows.

We discuss the effects of each of the optimization schedules on a synthetic dataset in the next section.

\section{Experiments}
\label{sec: exp}
We apply the proposed framework (Equations \ref{eq:similarity_matching}, \ref{eq:ksm-sc}, \ref{eq: ksm-manifold-scaled}) on synthetic and real datasets (high dimension), to study the optimization process and visualize the implicitly learned generative maps from the estimated latent representations and the kernel structure of the latent representations. \rev{We begin our discussion by examining objective in \Eqref{eq:ksm-sc} -- imposing sparsity constraints on the latent space -- on the synthetic dataset and real datasets in the following two sub-sections}.


\subsection{Applying KSM on simulated data with known latents}
\label{sec: exp-synthetic}

\textbf{Setup and Initialization:} To generate the synthetic data, we sample the dictionary $\mA \in \sR^{N\times K}, (N = 30, K = 16)$ from $\gN\left(0, \mI\right)$. The latent representations or \textit{sparse codes} were generated with a sparsity of $12.5\%$ with amplitude values ranging uniformly between $[4, 6]$,  generating $1500$ samples for this dataset.  Thereafter, we set $\mP_i = \vz_i\vz_i^T + \omega \mI,  \,\forall\,\in \{1, 2, \dots T\}$. 

We warm start the optimization process by initialization $\mW$ by injecting additive white Gaussian noise to the true dictionary $\mA$ value of standard deviation 0.3 \citep{ba2020deeply}, 
and initialize $\mM_i = \frac{\mW\mW^T}{2}, \,\forall\, i \in \{1, 2, \dots\}$, with the representations $\mZ$ initialized by performing sparse coding with $\mW$ as the dictionary. The values of the hyperparameters for the optimization process are outlined in Table \ref{tab:my-table}. Here $T_z$ refers to the number of gradient based updates to perform the \textit{Z-Step} in Algorithm \ref{algo:admm_sep}. To simplify the tuning process for $\omega$ we normalize $\mW$ to have unit norm rows and set $\omega=10^{-4}$. 

\begin{figure}[h!]
    \centering
    \includegraphics[width=\textwidth]{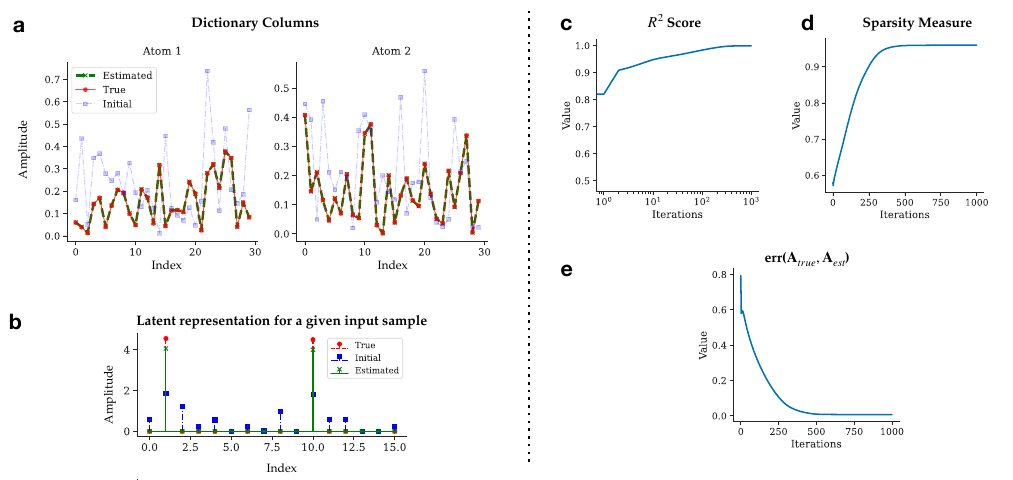}
    \caption{(a) Sample estimated dictionaries (generative maps) learned from the latent kernel structure (we show initial values, before optimization, the final values and the ground truth) (b) Sample estimated latent representation showing alignment with ground truth. (c) $R^2$ value of the predictions, (d) 1-Sparsity Measure (Appendix \ref{app: p-sparsity}) of the latents, and (e) the average similarity error between the estimated and the true generative maps over the optimization process.}
    \label{fig:synthetic_data_sc}
\end{figure}


\rev{\textbf{Discussion} \Figref{fig:synthetic_data_sc} shows the results from the optimization process performed under \textit{Schedule 3} on the synthetic data which highlight a couple of observations. First, \Figref{fig:synthetic_data_sc} (a) shows the true, initial and final estimated generative maps from the learned representations for two random atoms with the true and final maps showing alignment. Similar alignment can be observed for the learned representations in \Figref{fig:synthetic_data_sc} (b), which shows the representation vector for a random sample. Second, the learned representations show increasing level of sparsity (as defined in Appendix \ref{app: p-sparsity}) as shown in \Figref{fig:synthetic_data_sc}(d) as the optimization proceeds a consequence of the sparsity constraint imposed by the $l_1$ norm. Third, to ascertain the correctness of the estimated generative maps, we define the error metric as a sine similarity measure between the true and estimated atoms which is given by:
\begin{equation}
    \begin{aligned}
    \text{err}(\rmA_{\text{true}}, \rmA_{\text{est}}) &= \sqrt{1 - \theta_{\text{err}}^2}\\
    \text{where, }\theta_{\text{err}} &=\frac{1}{K}\sum\limits_{k=1}^K \frac{\left|\rva^{(k)^T}_{\text{true}} \rva^{(k)}_{\text{est}}\right|}{\left\|\rva^{(k)}_{\text{true}}\right\|_2\left\|\rva^{(k)}_{\text{est}}\right\|_2} \\
    \end{aligned}
\end{equation}

We observe that the estimated generative maps using the latent kernel structure progressively align with the true generative maps as demonstrated by the decreasing values of the error metric in \Figref{fig:synthetic_data_sc} (e). Finally, the correctness of the predictions made by using the learned representations and the estimated generative map is captured by the increasing trend in the $R^2$ value as shown in \Figref{fig:synthetic_data_sc} (d).}

\begin{figure}[h]
    \centering
    \includegraphics[width=\textwidth]{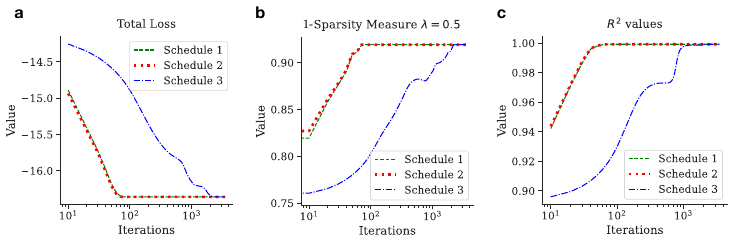}
    \caption{Comparison of the different optimization schedules on the synthetic dataset. The plot shows the convergence of the total loss, 1-sparsity measure (Appendix \ref{app: p-sparsity}) and $R^2$ values for the different schedules. We observe that \textit{Schedules 1} and \textit{2} converge at a similar rate, which is faster than \textit{Schedule 3}.}
    \label{fig: schedule-comparison}
\end{figure}

\textbf{Comparison of different schedules:} We also compare different schedules for the optimization process on the synthetic dataset. As alluded to in Section \ref{sec: computational_considerations}, the value of $\eta_Z$ under \textit{Schedule 3} is significantly smaller than the other proposed schedules. 
As a result, we expect \textit{Schedule 3} to converge at slower rate when compared the other schedules.  This can be observed in \Figref{fig: schedule-comparison} which shows the comparison of the different schedules on the synthetic dataset as described in this section. We see that both \textit{Schedules 1} and \textit{2} converge at a similar rate, which is faster than \textit{Schedule 3}. 

\subsection{Applying KSM on real datasets with unknown latents}
\textbf{Setup and Initialization:} We also applied the kernel matching framework to real datasets, namely, the \textit{MNIST dataset} ($N=784, K=500$) \citep{lecun1998gradient} and \textit{patches} extracted from Natural Scenes dataset ($N=256, K=192$) \citep{olshausen_emergence_1996} with significantly higher input dimensions. We extract $\sim$ 9000 samples total from 5 different classes in the MNIST dataset and around 40000 random patches ($16\times 16$) extracted from a total of 10 images (each of size $512\times512$) of the Natural Scenes dataset \citep{olshausen1996emergence}. Our approach involves using \textit{Schedule 1} of the optimization process on account of it showing faster rate of convergence in simulated experiments, as discussed in Section \ref{sec: computational_considerations}. We follow a similar initialization scheme for $\mW, \mM_i, \mP_i$ and $\mZ$ as discussed for the synthetic data experiments before. Table \ref{tab:my-table} contains the values of the hyperparameters used for the optimization process. The slower learning rate for $\mW$ indicates that the loss surface in higher dimension tends to be more complex with multiple local minima, flat regions and saddle points and consequently require that gradient based methods proceed at a slower rate to avoid instabilities, thereby requiring more iterations of the optimization algorithm.

\begin{figure}[h]
    \hspace*{-5mm}
    \includegraphics[scale=.78]{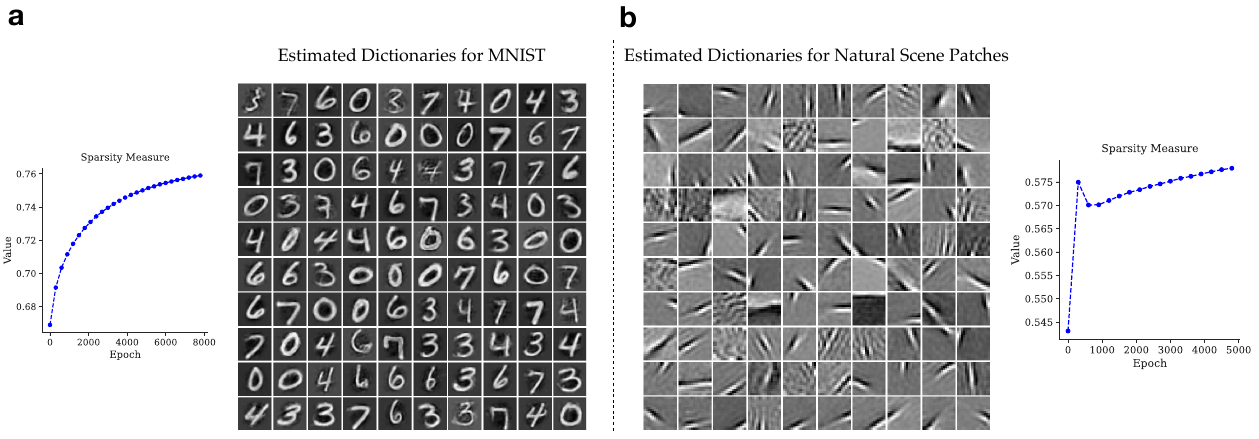}
    \caption{Estimated dictionaries using the implicit generative map on (a) the MNIST dataset (b) on patches extracted from natural scenes (100 atoms visualized) show that the learned representations implicitly learn generative maps that capture data features. The increasing sparsity measure also indicates that the learned representations progressively get sparser over the optimization process. Data mean normalized and whitened in pre-processing.}
    \label{fig:real_data_sc}
\end{figure}

\Figref{fig:real_data_sc} shows the dictionary or the generative maps learned from the representations on the MNIST and patches dataset. Each column of the dictionary represents a digit like structure for the MNIST dataset and gabor like structures for the patches extracted from Natural Scenes demonstrating selective receptive fields as discussed in \citep{olshausen_emergence_1996}. The sparsity of the learned representations progressively for both the datasets indicating increasing selectivity of the learned representations to the input data manifold. 

\FloatBarrier


\subsection{Synthetic data with Simplex Prior}
\label{sec: exp-synthetic-simplex}
We demonstrate the flexibility of the framework under different latent priors. 
We apply the objective in Proposition \ref{app: prop-ksm-manifold} to the half moons dataset \citep{pedregosa2011scikit} similar to approach undertaken in \citep{luther_kernel_2022}.
by applying the objective in Proposition \ref{prop: ksm-manifold} to a synthetic moons dataset \citep{pedregosa2011scikit}. We sample $N=2500$ points from the dataset with an added Gaussian noise of 0.01 in magnitude, learning $K=12$ receptive fields serving as anchor points that are used to piece-wise approximate the manifold curvature. To initialize the optimization process outlined by Algorithm \ref{algo:admm_sep}, we initialize $\mW$ using random points on the input manifold. The initial latent representations were drawn from a standard normal distribution and projected on to the probability simplex. For the Lagrange multipliers, we initialize $\mM_i = \frac{\mW\mW^T}{2}\,\forall\,i \in \{1, 2, \dots, T\}$.  The choice of hyperparameters in this experiment is also outlined in table \ref{tab:my-table}. Note, in this case we do not have the parameter $\lambda$ and regularization is simply done by the parameter $\omega$. 

\begin{figure}[h]
    \centering
    \includegraphics[width=\textwidth, trim={0, 0.5cm, 0, 0.1cm}, clip]{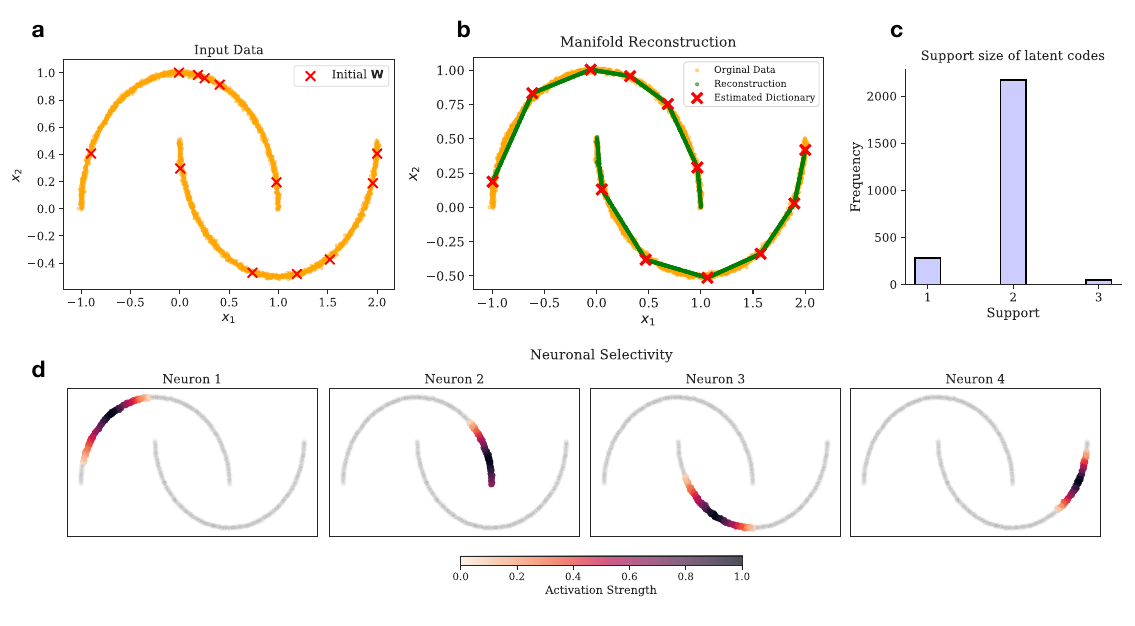}
    \caption{(a) An intrinsically one dimensional manifold represented by the two-moons dataset. (b) The piecewise approximated reconstruction by estimating the receptive fields through the implicit generative map learned from the representations. (c) $L_0$ norm of the latent codes indicating number of active neurons for each input sample. We see that most examples activate 2 neurons within a population indicating piece-wise linear reconstruction. This is directly tied to the intrinsic dimensionality of the input manifold which is one less than the modal $L_0$ norm. (d) Local selectivity of the neuronal population indicates that the neuronal activity is sensitive to local regions within the input manifold.}
    \label{fig:ksm-kds}
\end{figure}

\textbf{Discussion:}  \Figref{fig:ksm-kds} shows the results of the optimization process on the synthetic moons dataset. These estimated receptive fields, together with the learned representations, piece-wise linearize the input data manifold as shown in \Figref{fig:ksm-kds}(b). From \Figref{fig:ksm-kds}(c), we can note that a high percentage of the samples on the input manifold activate two neurons at a time in the population, since most of the learned representations have a \textbf{support of size (\# of nonzero elements) of 2}, suggesting the input data manifold can be piecewise-linearized using one dimensional line segments. This can be used to estimate the intrinsic dimension of the manifold, which in the case of the moons dataset is one, thereby suggesting that \textit{the intrinsic dimension of the manifold is one less than the mode of the support of the latent representations}. From \Figref{fig:ksm-kds}(d), we observe four sample neurons out of $K=12$ that are activated by points on the input data manifold. The tuning curves of these neurons tile the manifold by having the maximum activations at the receptive fields indicated by the red crosses in \Figref{fig:ksm-kds}(b), with the activation strength spreading along the geometry of the input data manifold. This suggests that metabolic constraints on neural activity tunes the learned representations to capture the geometric structure of the input data manifold.

\begin{table}
    \centering
    \caption{Choice of hyperparameters for the optimization process}
    \label{tab:my-table}
    \begin{tblr}{
      cells = {c},
      row{1} = {Silver},
      cell{1}{1} = {r=2}{},
      cell{1}{2} = {c=3}{},
      vlines,
      hline{1,3,11-12} = {-}{},
      hline{2} = {2-5}{},
    }
    \textbf{Parameter} & {\textbf{Sparse Dictionary }\\\textbf{Learning}} &                &                  & {\textbf{Manifold}\\\textbf{Learning}} \\
                       & \textbf{Synthetic}                               & \textbf{MNIST} & \textbf{Patches} & \textbf{Moons}                         \\
    $K$                & 16                                               & 500            & 192              & 12                                     \\
    $\lambda$          & 0.5                                              & 0.1            & 0.9              & --                                      \\
    $\omega$           & $10^{-4}$                                        & $10^{-4}$      & $10^{-4}$        & 0.1                                    \\
    $\rho$             & 1.0                                              & 1.0            & 1.0              & 1.0                                    \\
    $\eta_W$           & 0.01                                             & $10^{-4}$      & $10^{-4}$        & $1.5\times10^{-3}$                     \\
    $T_z$              & 15                                               & 15             & 15               & 15                                     \\
    batch size $(B)$   & 128                                              & 128            & 128              & 128                                    \\
    max. epochs        & 3500                                             & 8000           & 5000             & 5500                                   \\
    $R^2$ value        & 1.0                                              & 0.78           & 0.86             & 0.98                                   
    \end{tblr}
\end{table}

\FloatBarrier

\begin{figure}
    \centering
    \includegraphics[scale=0.65, trim={0, 0, 3.3cm, 0}, clip]{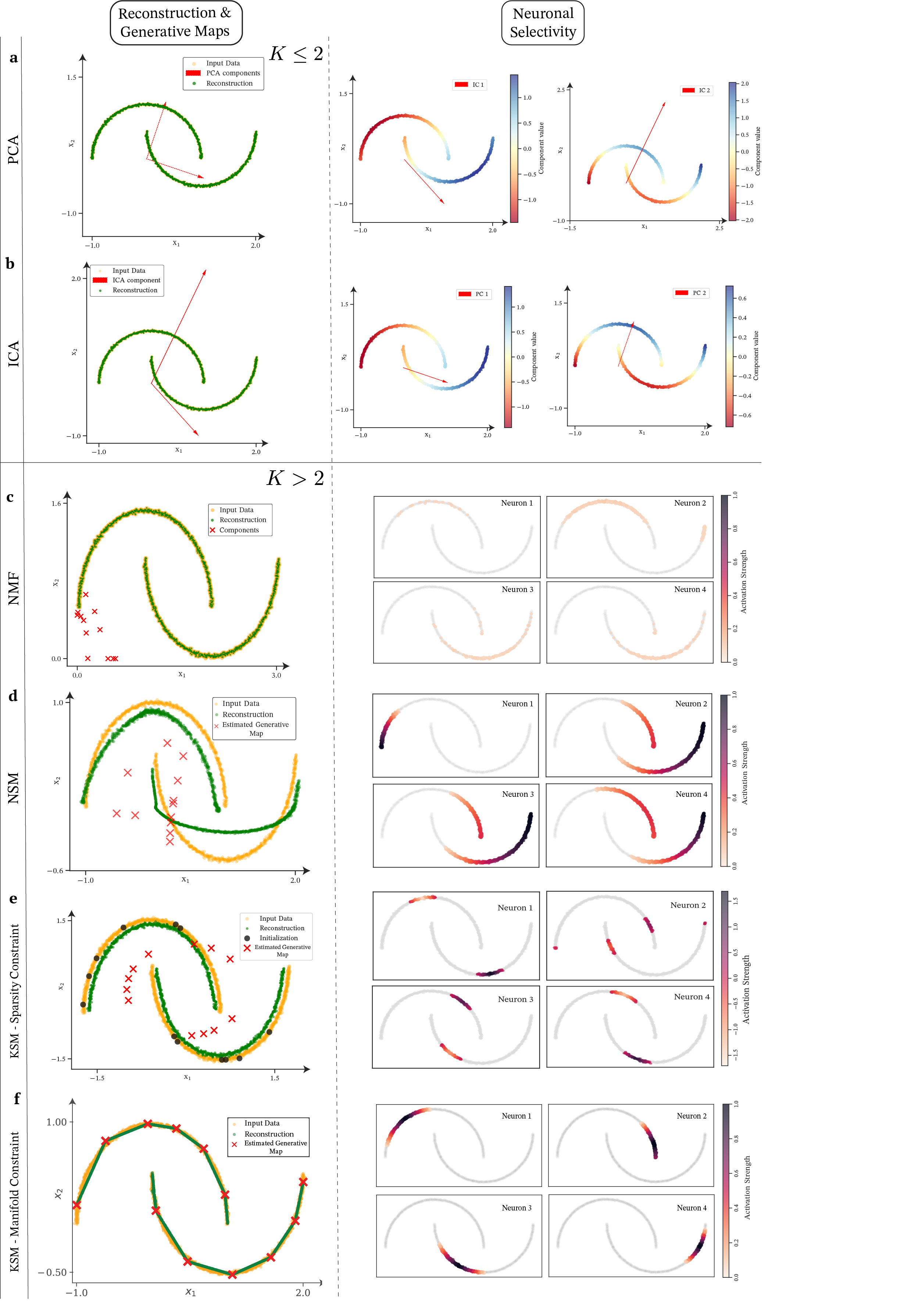}
    \caption{\rev{(a)-(f) Estimated Generative Maps alongside predictions (left) and neuronal selectivity (right) for different kernel-based approaches for representation learning on the moons (2d) dataset. (a), (b) show undercomplete representation learning approaches in the form of PCA and ICA. (c)-(f) Overcomplete representation learning showing locally selective representations. (e)-(f) We see that how the latent structure affects the learned maps, with manifold constraints (simplex prior) yielding locally interpretable generative maps that are representative of the input. }}
    \label{fig:kernel-comparison}
\end{figure}

\rev{
\subsection{Comparison with other kernel-based representation learning methods}
\label{subsec: comparison}

To put things into perspective, we compare our proposed approach with other representation learning approaches employing kernel matching methods as discussed in \S \ref{sec:intro}. We focus on both methods learning undercomplete representations such as PCA \citep{pehlevan_why_2018} and ICA \citep{bahroun2021normative} as well as overcomplete representations such as non-negative matrix factorization (NMF) \citep{lipshutz2023normative,pehlevan2014hebbian} and non-negative similarity matching (NSM) \citep{sengupta_manifold-tiling_2018}. We build intuition through \Figref{fig:kernel-comparison} which shows the prediction and the neuronal selectivity for different methods on the 2d moons dataset and then focus on the different generative models on a higher dimensional real dataset where undercomplete representations are typically useful (\Figref{fig:real-data-pca-ica}). We limit the discussion to PCA and ICA in this case. 
}

\vspace{1em}

\rev{

\textbf{Overcomplete, constraint-driven kernel matching helps learn nonlinear manifolds with local selectivity:}\\
PCA and ICA, while learn a generative map, are limited to undercomplete representations and as such learn a linear subspace. Consequently, the maps learned in this case represent global features of the data (directions of maximum variance -- for PCA-- or gaussianity -- for ICA) and are not particulary represented of local structures of the manifold. This can be seen by the lack of selectivity for individual neurons (neurons respond to the entire manifold). In contrast, overcomplete representations, can be locally selective (\Figref{fig:kernel-comparison}, $K>2$). \\

\textbf{Structural constraints on the latent space help learn locally interpretable generative maps:}\\
From \Figref{fig:kernel-comparison}, we see that methods such as non-negative factorization (for overcomplete representations) alongside PCA, ICA for undercomplete representations are able to predict the input data well, however, the learned generative maps tend to be non-localized, either capturing global features of the data (PCA, ICA) or being difficult to interpret (NMF). Furthermore, the lack of an inherent generative structure of the NSM model, while making it suitable for learning locally selective representations, does not allow us to decode the said representations back to the input space thereby missing a layer of interpretability. On the other hand, the proposed KSM framework with structural constraints on the latent space is able to learn locally selective representations while being able to decode the representations back to the input space (\Figref{fig:kernel-comparison} (e) and (f)). Furthermore, by \textit{adjusting the constraints on the kernel structure} and \textit{the latent representations}, we are able to learn locally interpretable generative maps, representative of the input manifold, exhibiting receptive-field like characteristics for the local representations (\Figref{fig:kernel-comparison}(f)).

\begin{figure}[h]
    \centering
    \includegraphics[scale=1.2, trim={0.16cm, 0.17cm, .18cm, .6cm}, clip]{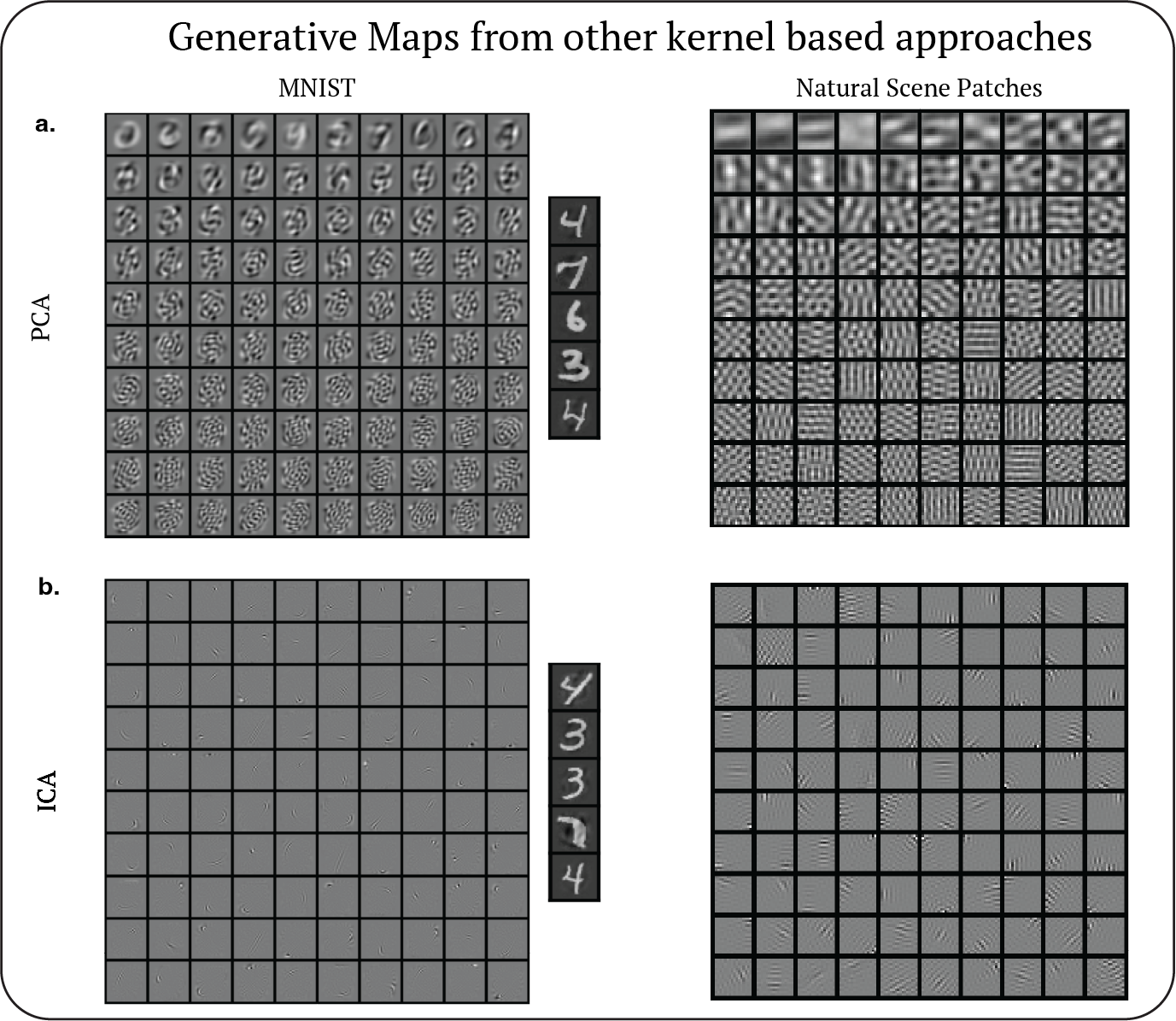}
    \caption{\rev{Estimated generative maps from PCA and ICA. Both rely on kernel matching objectives, however, the learned generative maps are non-localized, capturing global features of the data. Data whitened and mean standardized in pre-processing.}}
    \label{fig:real-data-pca-ica}
\end{figure}

We extend this intuition to higher dimensional datasets (MNIST and Natural Scenes), where we compare the generative maps learned by PCA and ICA alongside sparsity constraints on the latent space through KSM (\Figref{fig:real_data_sc}, \Figref{fig:real-data-pca-ica}). Here, we see that the generative maps learned by PCA and ICA tend to be non-localized, capturing global directions of variations within the data (basic strokes and edges). In contrast, the KSM framework learns locally interpretable generative map, capturing digit like filters for MNIST and gabor like filters for patches extracted from Natural Scenes, which are known to be good representations of the data \citep{olshausen_emergence_1996}, with higher $r^2$ values for the predictions.

}

\FloatBarrier

\section{Towards a biologically plausible implementation (SMPC architecture)}
\label{sec:bio-plausible}

As mentioned before, kernel matching frameworks have been used to propose a biologically plausible model for learning representations \citep{pehlevan_why_2018,luther_kernel_2022}. These models learn representations by minimizing the distance between the input and latent kernel using feedforward and lateral connections which are learned using Hebbian and Anti-Hebbian rules. However, while good at learning representations representing the input, there is no explicit mechanism to learn generative maps in such frameworks. We conjecture that learning the kernel structure in the representation space could provide one way to extend kernel matching approaches to learn generative maps, which traditionally require top-down interactions i.e. feedback from higher cortical layers to lower ones. This idea of top-down interactions between different cortical layers forms the basis of predicting coding approaches which have been used to provide a mathematical framework for visual processing in the brain \citep{rao_predictive_1999,keller_predictive_2018,boutin2020effect,boutin_sparse_2021}. Using the objective developed in Proposition \ref{app: prop-ksm-obj}, we discuss a biologically plausible extension to the classical kernel matching framework in \citep{pehlevan_why_2018,luther_kernel_2022} incorporating top-down interactions for learning generative maps alongside representations.

We start with articulating a relaxed version of the objective in Equation \ref{eq:ksm-sc} given as follows:
\begin{align}
    \label{eq:relaxed_ksm}
    \underset{\mY, \mZ, \mOmega}{\min}\,\,& -\frac{1}{2T^2}\Tr(\mX^T\mX\mY^T\mY) + \frac{\lambda}{T}\|\mZ\|_{1,1} + \frac{\rho}{T}\|\mY - \mOmega\mZ\|_F^2
\end{align}
We introduce a standard kernel using an intermediate representation $\mY$ in the trace term in \Eqref{eq:relaxed_ksm}, replacing the structured latent kernel in \Eqref{eq:ksm-sc}. This intermediate representation is related to the final latents $\mZ$ through a linear transformation $\mOmega$, which is enforced by the penalty term $\frac{\rho}{2T}\|\mY - \mOmega\mZ\|_F^2$ in the objective, where $\rho$ controls the strength of the penalty. Under \rev{high} penalty values, i.e, when $\mY \approx \mOmega\mZ$, the trace term in the objective reduces to computing a similarity measure between the standard kernel of the input matrix $\mX$ and the structured kernel of the representations $\mZ$, with the structure determined by the matrix $\mOmega^T\mOmega$. Unlike the formulation in \Eqref{eq:ksm-sc}, we relax the constraints on the kernel structure $\mOmega$ for the formulation above, allowing it to be governed by the data itself. Accordingly, as outlined in \S \ref{sec:ksm-implicit}, the implicit generative map can be articulated as $\hat\mA = \dfrac{\mX\mZ^T(\mOmega^T\mOmega)}{T}$, where $\mOmega^T\mOmega$ determines the kernel structure in the latent space, similar to $\mH^{-1}$.

\textbf{Connection to classical MDS and stabilizing the learning dynamics:}
We augment the objective in \Eqref{eq:relaxed_ksm} by adding a penalty term comprising of the Frobenius norm of the kernel of $\mY$, resulting in the update objective as follows:
\begin{equation}
    \label{eq:relaxed-ksm-upper-bound}
    \begin{aligned}
        \undermin{\mY, \mZ, \mOmega}& -\frac{1}{2T^2}\Tr(\mX^T\mX\mY^T\mY) + \frac{\alpha}{4T^2}\Tr(\mY^T\mY\mY^T\mY) + \frac{\lambda}{T}\|\mZ\|_{1,1} + \frac{\rho}{T}\|\mY - \mOmega\mZ\|_F^2 \\
        \Longleftrightarrow \undermin{{\mY, \mZ, \mOmega}}&\,\,\frac{1}{4T^2}\|\mX^T\mX - \mY^T\mY\|_F^2 + \frac{\lambda}{T}\|\mZ\|_{1,1} + \frac{\rho}{T}\|\mY - \mOmega\mZ\|_F^2
    \end{aligned}
\end{equation}
where, the lower objective is obtained by completing the squares. For our case we choose $\alpha=1$. The addition of the term $\frac{1}{4T^2}\Tr(\mY^T\mY\mY^T\mY)$ to the objective  in \eqref{eq:relaxed-ksm-upper-bound} introduces a penalty on the norm of the intermediate kernel which is absent in \Eqref{eq:relaxed_ksm}. This addition transforms the objective in \Eqref{eq:relaxed-ksm-upper-bound} into an \textit{upper bound} on the objective in \Eqref{eq:relaxed_ksm}. The kernel norm penalty enforces a stronger similarity between the input and the intermediate representations by aligning the norm of the respective kernels. Furthermore, this converts the trace term in \Eqref{eq:relaxed_ksm} into a Classical Multidimensional Scaling (MDS) objective \citep{borg2007modern}. A key benefit of this augmentation is the stabilization of the learning dynamics during optimization with respect to the intermediate representation $\mY$. The original formulation in \Eqref{eq:relaxed_ksm} can result in runaway dynamics in certain directions of the optimization space, due to the strong concavity introduced by the negative of the trace term with respect to $\mY$. Incorporating the kernel norm penalty address this issue by bounding the objective in \Eqref{eq:relaxed-ksm-upper-bound} from below. Consequently, the entire objective now comprises of two parts -- a \textit{similarity matching term} between the input $\mX$ and the intermediate representations $\mY$ and prediction error term between the intermediate and the final representations $\mY$ and $\mZ$ respectively. Employing Legendre transformation on the objective in \Eqref{eq:relaxed-ksm-upper-bound} allows us to derive an online version of the objective similar to the ones as discussed in \S \ref{sec: similarity_matching}. We articulate the online objective in the proposition below:


\begin{proposition}
    \label{prop: smpc-online}
    Let $R_2$ be the optimization problem as described below. Then $R_2$ is equivalent to \Eqref{eq:relaxed-ksm-upper-bound}.
    \begin{equation}
        \label{eq:smpc}
        \begin{aligned}
        R_2: \undermin{\mY, \mZ, \mOmega, \mW}\, \undermax{\mM} &\,\frac{1}{T}\sum\limits_{i=1}^T \left(-\vx_i^T\mW^T\vy_i + \frac{1}{2}\Tr(\mW^T\mW) + \frac{1}{2}\vy_i^T\mM\vy_i -\frac{1}{4}\Tr(\mM^T\mM)+  \lambda\|\vz_i\|_1 + \frac{\rho}{2}\|\vy_i - \mOmega\vz_i\|^2\right)\\
        \end{aligned}
    \end{equation}
\end{proposition}

We note that the objective described in \Eqref{eq:smpc} is a min max optimization problem. We exchange the order of optimization with respect to optimization variables $\mM$ and the latent representations $\mZ$ and $\mY$. The first exchange between $\mM$ and $\mZ$ is possible on the account of the objective being decoupled with respect to these variables. The second exchange between $\mM$ and $\mY$ can be done as the objective satisfies the saddle point property (proof in Appendix, similar to \citep{pehlevan_why_2018}). The resultant optimization problem can be written as:
\begin{proposition}
    \label{prop: smpc-minmax}
    Let $R_3$ be the optimization problem as described below. Then $R_3$ is equivalent to \Eqref{eq:smpc}.
    \begin{equation}
        \label{eq:smpc-minmax}
        \begin{aligned}
        R_3: \undermin{\mOmega, \mW}\,\undermax{\mM}\,\undermin{\mZ, \mY} &\,\frac{1}{T}\sum\limits_{i=1}^T \left(-\vx_i^T\mW^T\vy_i + \frac{1}{2}\Tr(\mW^T\mW) + \frac{1}{2}\vy_i^T\mM\vy_i -\frac{1}{4}\Tr(\mM^T\mM)+  \lambda\|\vz_i\|_1 + \frac{\rho}{2}\|\vy_i - \mOmega\vz_i\|^2\right)\\
        \end{aligned}
    \end{equation}
\end{proposition}

\begin{figure}[h]
    \centering
    \hspace*{-4mm}
    \includegraphics[scale=0.65]{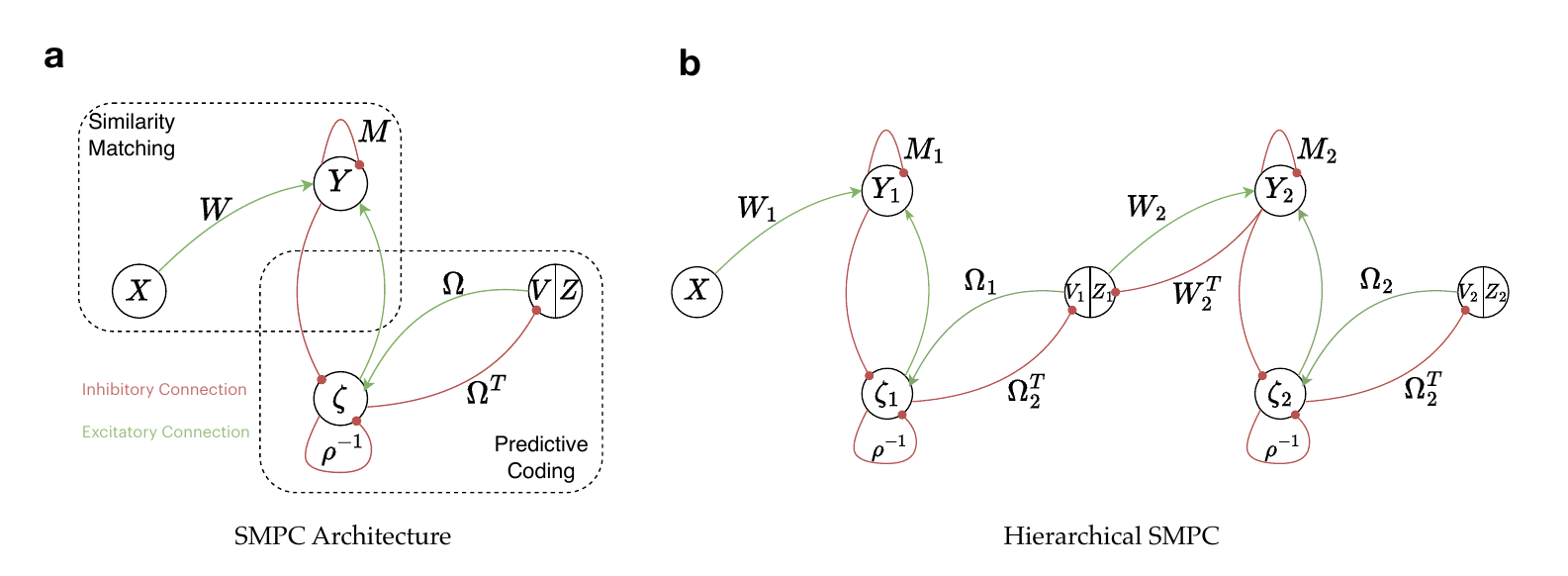}
    \caption{(a) The SMPC architecture for kernel similarity matching.  The bottom-up interactions in encoding $\mY$ from the input $\mX$ and the top-down interactions in predicting $\mY$ from the latent activity $\mZ$ are modulated through the interneuron $\bm\zeta$. The inhibitory connections and the excitatory connections are obtained from the gradient dynamics of the optimization process (Appendix \ref{app: smpc}) (b) Hierarchical extension of the SMPC architecture, where the learned latents in the first layer act as an input in the second layer. Unlike classical feed-forward models for representation learning, there is an feedback loop from the higher layer to the lower layer. In our implementation, the neuronal population denoted by $\mY$ acts as an inhibitory neuron and the population denoted by $\mV|\mZ$ acts as an excitatory neuron in accordance with \textit{Dale's law}, however, the interneuron sends out mixed signals propagating the excitation from the higher layers and inhibition from the lower layers. The interneuron population could be split to make the architecture satisfy Dale's law, however, we omit that here for simplicity.}
    \label{fig: smpc-architecture}
\end{figure}

We can now optimize the objective in \Eqref{eq:smpc-minmax} by applying stochastic gradient descent-ascent algorithm, first optimizing with respect to the representations $\mY$ and $\mZ$ followed by updating the sample independent parameters $\mW, \mM, \mOmega$. \rev{We articulate the gradient dynamics of the optimization process in Algorithm \ref{algo:smpc-bioplausible} similar to \cite{pehlevan_why_2018}} and use the dynamics of each of the variables to derive a predictive coding network (\Figref{fig: smpc-architecture}) as discussed in \citep{bogacz_tutorial_2017}, where the synaptic weights are learned using Hebbian and anti-Hebbian rules. Here $\mY$ and $\mZ$ model neuron activations, with $\mZ$ being modeled as a two cell neuron, whereas, the sample independent terms $\mW, \mM$ and $\mOmega$ model the synapses. 

{
\begin{algorithm}
    \color{black}
    \caption{Optimization of the SMPC objective in \Eqref{eq:smpc-minmax} using Hebbian updates}
    \KwIn{data matrix $\mX = [\vx_1, \vx_2, \dots, \vx_T],\ \vx_t \in \sR^{N}$}
    \KwOut{latent matrix $\mZ = [\vz_1, \vz_2, \dots, \vz_T],\ \vz_t \in \sR^{K}$ }
    \textbf{Intermediate}: latent matrix $\mY = [\vy_1, \vy_2, \dots, \vy_T],\ \vy_t \in \sR^{K}$\\[2pt]
    \BlankLine

    \textbf{(1)} At $t=0$, initialize the synaptic weight matrices $\mW_1,\ \mM_1,\ \mOmega_1$. $\mM_1$ must be symmetric positive definite. (Appendix \ref{app: smpc})\;

    \For{\textbf{(2)} $t = 1, 2, \dots, T$}{
        For every input $\vx_t$\;

        \textbf{(3)} Neural Dynamics for iteration $\gamma$: repeat until convergence\;
        $\dfrac{d\vy_t(\gamma)}{d\gamma} \gets \mW\vx_t - \mM \vy_t + \bm{\zeta}_t$\;
        $\tau\dfrac{d\vv_t(\gamma)}{d\gamma} = \dfrac{\vz_t - \vv_t}{\eta} - \mOmega^T\bm{\zeta}_t$;\quad $\vz_t(\gamma) = f(\vv_t(\gamma))$\;
        $\dfrac{d\bm{\zeta}_t(\gamma)}{d\gamma} = \mOmega\vz_t - \vy_t - \rho^{-1}\bm{\zeta}_t$\;

        \textbf{(4)} Synaptic Plasticity \tcp*[r]{Hebbian updates}
        $\mOmega_{t+1} = \mOmega_t - \eta_\mOmega \left(\bm{\zeta}_t\vz_t^T\right)$\;
        $\mW_{t+1} = \mW_t + \eta_\mW \left(\vy_t\vx_t^T - \mW_t\right)$\;
        $\mM_{t+1} = \mM_t + \eta_\mM \left(\vy_t\vy_t^T - \mM_t\right)$\;
    }

    \label{algo:smpc-bioplausible}
\end{algorithm}
}
    
\rev{Algorithm \ref{algo:smpc-bioplausible} outlines the steps of the optimization process, with $f$ denoting the nonlinear activation based on the constraints on the latent space. We formulate the discretized version of the algorithm in Appendix \ref{app: smpc}, which is used in our numerical experiments. The interactions between the intermediate latent $\mY$ and the final latent representation $\mZ$ are modeled using an inter-neuron $\bm{\zeta}$, which mediates the bottom-up and top-down interactions as discussed below.}

\textbf{Modeling the prediction error using interneurons:} The prediction error between the final latent representation $\mZ$ and the intermediate representation $\mY$ is modeled using interneurons denoted by $\bm{\zeta}$. Interneurons have been used in earlier work to describe how neurons mediate recurrent communication to perform operations such as statistical whitening \citep{lipshutz2022interneurons}. These interneurons often called error neurons in predictive coding literature, are used to minimize prediction errors in line with redundancy reduction hypothesis of neural coding, allowing the brain to focus on unexpected stimuli to improve the accuracy of predictions \citep{simoncelli2001natural, friston2005theory}. In our case, these interneurons mediate between bottom-up interactions stemming from encoding $\mY$ through kernel similarity matching with the input $\mX$ and the top-down interactions in predicting $\mY$ from the latent  representations $\mZ$. As a result, we call the architecture denoted in \Figref{fig: smpc-architecture}(a) \textit{as the SMPC (similarity matching predictive coding) architecture}, where the learning dynamics of the synapses follow local plasticity rules. Furthermore, the structure of the SMPC architecture can be extended to form a hierarchical network which can be used to model information flow between different cortical layers (\Figref{fig: smpc-architecture} (b)). We keep the discussion to a single layer network and leave the extension to a hierarchical network as future work.


\FloatBarrier
\subsection{Numerical Experiments}

We now investigate the ability of the SMPC architecture to implicitly learn generative maps on simulated datasets in the section below. We begin with sparsity constraints on the latent space. 

\begin{figure}[h]
    \centering
    \includegraphics[width=\textwidth]{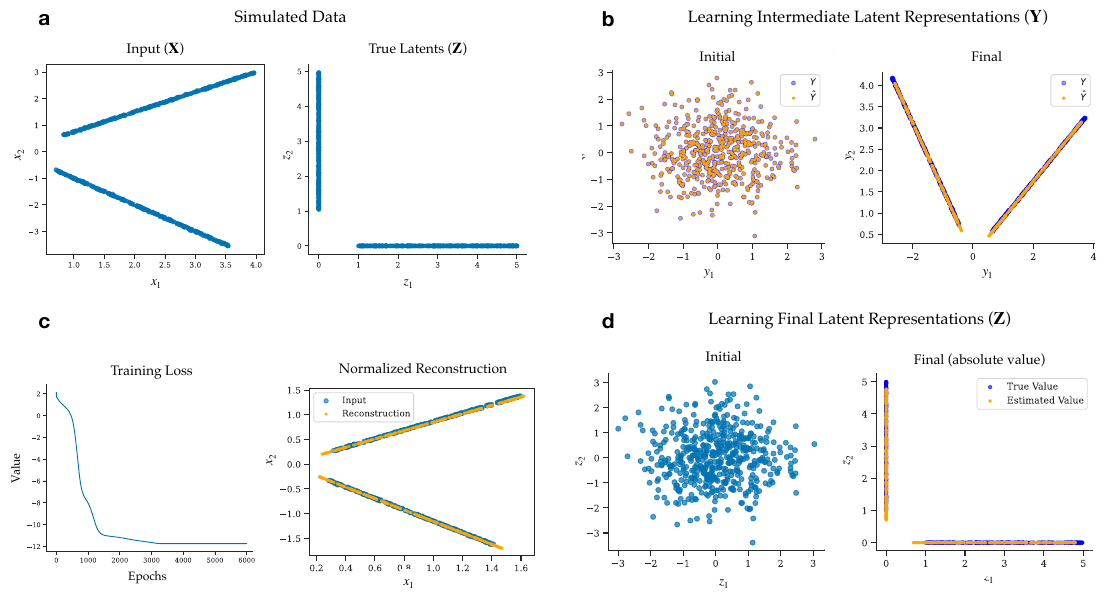}
    \caption{Results of applying the SMPC architecture on a 2D dataset. (a) Simulated 2D dataset with true latent representations $\mZ$. (b) Intermediate representations ($\mY$) governed by bottom-up interactions from the input $\mX$ and their predictions from top-down interactions ($\hat\mY = \mOmega \mZ$). The interneuron $\bm{\zeta}$ effectively modulates the communication between the intermediate and final latents. (c) Decrease in the overall objective value as the optimization proceeds along with the predicted derived from the \textit{implicit generative map} using the learned representations $\mZ$ (shown in (d)) and the synaptic weights $\mOmega$. We show normalized input and prediction, normalized by data variance, to make sure any scaling effects in reconstruction are minimized. (d) The final representations $\mZ$ learned by the SMPC architecture with the initial values prior to the optimization showed on the left.}
    \label{fig:2d-smpc-results}
\end{figure}

\textbf{Setup:} We applied the SMPC architecture (\Figref{fig: smpc-architecture}(a)) to simple 2D dataset in \Figref{fig:2d-smpc-results}(a).  The dataset consists of $N = 512$ points generated by atoms of dictionary $\mA \in \sR^{2\times 2}$, where the atoms are chosen along two directions such that correlation between the atoms is $0.14$. This low mutual coherence between the atoms is necessary for sparse dictionary learning \citep{donoho2003optimally, donoho2006compressed}, \rev{and increasing the coherence led to a decrease in the reconstruction performance}. The latent representations $\mZ$ are generated along the $x$ and $y$ axis as shown in \Figref{fig:2d-smpc-results}(a). Additionally, a small amount of additive Gaussian noise (standard deviation = 0.01) was added to the  samples. 

To speed up the optimization process, the synaptic updates were performed using batches sampled from the data. For this 2d dataset, the batch size was set to be equal to the length of the dataset. The learning rates for the neuronal dynamics $\eta_\mY, \eta_\mZ$ were set to $0.1$ and run for $10$ iterations each, with $\eta_\mW = 0.05 \ll \eta_\mM = 0.1$ as suggested in \citep{luther_kernel_2022} and are run for $5$ iterations to indicate the slower learning cycle of the synaptic weights. Our implementation assumes that interneurons $\zeta$ are highly sensitive with the highest learning rates, implying that their dynamics remain at fixed points throughout the optimization process. The values of $\rho$ and $\lambda$ were set to $5.$ and $0.8$ respectively. The initialization details for the optimization process are provided in the appendix. The results of the simulation are shown in \Figref{fig:2d-smpc-results}.

\textbf{Discussion:} Our results indicate that the model successfully recovers both the latent structure and its corresponding values (\Figref{fig:2d-smpc-results}(d)). The interneuron facilitates  bottom-up and top-down communication between the intermediate and the final latents given by $\mY$ and $\mZ$ respectively. This is evident in \Figref{fig:2d-smpc-results} (b) where the intermediate latent $\mY$ aligns well with its corresponding top-down prediction, $\hat\mY = \mOmega \mZ$. To compute the predicted input, we use $\hat\mA = \frac{\mX\mZ^T(\mOmega^T\mOmega)}{T}$ to compute the generative map, as discussed in the previous section and articulated in \S \ref{sec:ksm-implicit}, with the prediction being given as $\hat\mX = \frac{\mX\gK_\mZ}{T}, \,\gK_\mZ = (\mOmega\mZ)^T(\mOmega\mZ)$. Here $\mZ$ and $\mOmega$ were estimated from the optimization process. While $\hat\mX$ as calculated before, captures the structure of the input data $\mX$, in our experiments, we found that the predicted input $\hat\mX$ aligns with the input data $\mX$ when the data is variance normalized, i.e. $\vx_i \gets \dfrac{\vx_i}{\sqrt{\frac{1}{T}\sum\limits_{j=1}^T\|\vx_j\|^2_2}}$. This is demonstrated in \Figref{fig:2d-smpc-results}(c). A reason for this could be the fact that we are matching kernels instead of explicit reconstruction, suggesting that while correlation structure is preserved, the actual data norm may not be. 

\begin{figure}[h]
    \hspace{-5mm}
    \includegraphics[scale=1.25]{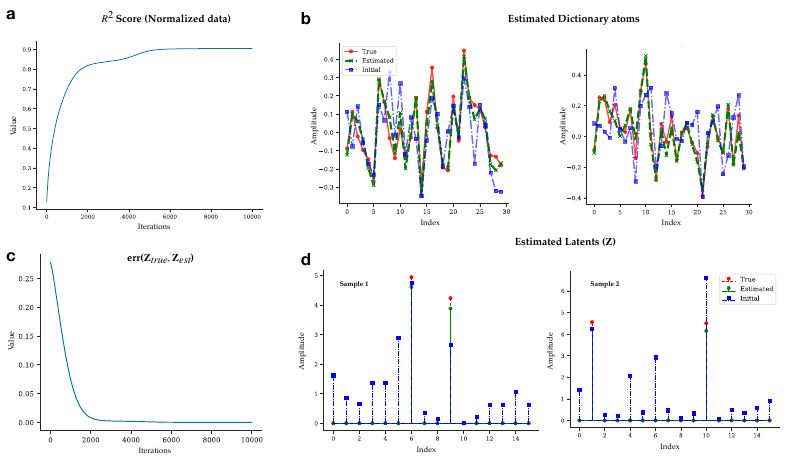}
    \caption{Applying the SMPC architecture on a higher dimensional dataset ($N=30$). (a) $R^2$ score plotted for normalized input ($\mX$) and predicted input $(\hat\mX$). (b) The \textit{implicit generative maps} computed using learned representations ($\mZ$) (shown in (d)) and the synaptic weights $\mOmega$. (c) Sine similarity error between the true and estimated latents ($\mZ$) with (d) showing two sample latents -- the true value, the initial value before optimization and the final value after optimization. The estimated latents align with the true latents, however, we see the lack of constraints on the kernel structure affects the $R^2$ value ( < 1) which reduces when the mutual coherence in the dictionary columns increase.}
    \label{fig: smpc-30-results}
\end{figure}

\textbf{Scaling up the data dimension}: Next, we scaled up the data dimensions to $N=30$ (similar to \S \ref{sec: exp-synthetic}) to understand the behavior of the SMPC architecture on higher dimensional dataset. We generate the data using a similar process as before (\S \ref{sec: exp-synthetic}), except the dictionary $\mA$ was subjected to an additional Gram-Schmidt orthogonalization to ensure that the mutual coherence between the atoms was low. The amplitude latent values, $\mZ$ were sampled independently from a $\gN(5, 1)$ distribution, with $87.5\%$ of the positions along a given dimension for every latent set to zero to ensure sparse representations. This process is similar to the generated latents shown in \Figref{fig:2d-smpc-results}(a). The initialization details for the implementation are discussed in the appendix. The results of the optimization process are shown in \Figref{fig: smpc-30-results}. \Figref{fig: smpc-30-results}(a) shows the $R^2$ value computed for the normalized input and the normalized predictions from the \textit{implicit generative maps}, computed as before, and the estimated latents ($\mZ$). We see that the model is able to provide good predictions (reflected by the $R^2$ value of 0.9). \Figref{fig: smpc-30-results}(b) shows the estimates of the implicit generative maps ($\hat\mA$). \Figref{fig: smpc-30-results}(c) shows the sine similarity between the estimated latents and the true values which is computed as $\displaystyle\text{err}(\mZ_{true}, \mZ_{est}) = \dfrac{1}{T}\sum\limits_{i=1}^T\left[1 - \left(\dfrac{\langle\vz_i^{(t)}, \vz_i^{(e)}\rangle}{\|\vz_i^{(t)}\|\|\vz_i^{(e)}\|}\right)^2\right]$, where $\vz_i^{(t)}$ are the true latent values and $\vz_i^{(e)}$ are the estimated latent representations. The error metric goes to zero as the estimated latents align with the true latents as shown in \Figref{fig: smpc-30-results}(c). \Figref{fig: smpc-30-results}(d) shows the true values, initial values prior to optimization and the estimated values post optimization. We observe that the estimated latents align with the true latents, however, the relaxed nature of the optimization problem affects the $R^2$ value which is less than 1, which suggests that imposing the constraints on the latent kernel structure plays a role in improving the implicit generative maps. Finally, varying $\alpha$ in \Eqref{eq:relaxed-ksm-upper-bound}, we found that $\alpha=1$ produced the best $R^2$ value.

\FloatBarrier

\textbf{Imposing metabolic constraints capture the curvature of the data manifold, but proper prediction warrants appropriate kernel structure:}
Finally, we apply metabolic constraints on the latent activity, imposed in the form of a probability simplex, to understand how the representations capture information about the input geometry in the SMPC architecture (similar to \S \ref{sec: exp-synthetic-simplex}). The initialization details for the optimization process are discussed in the appendix, with \Figref{fig: smpc-manifold-results} demonstrating the results at the end of the  optimization process. \Figref{fig: smpc-manifold-results} (a) demonstrates the input data manifold (we use the moons dataset \citep{pedregosa2011scikit} similar to \S \ref{sec: exp-synthetic-simplex}), alongside the predicted values at the beginning of the optimization process and the final predicted values. Additionally, we plot the implicit generative maps, estimated using the circuit parameters (red cross). We see that the predicted inputs are able to capture the curvature of the input manifold and correlate well (good similarity measure) with the input data, with the implicit generative maps following the curvature of the individual manifolds to form. However, the prediction does not align perfectly with the input due to the relaxed nature of the optimization problem, which lacks any constraints on the kernel structure of the latent space.  Nevertheless, we see emergence of local receptive field properties of individual neurons (shown in \Figref{fig: smpc-manifold-results} (b)) similar to the ones observed in \S \ref{sec: exp-synthetic-simplex}, suggesting that the simplex constraints tune neurons to capture the geometry of the input manifolds.



\begin{figure}[h]
    \hspace{-5mm}
    \includegraphics[width=\textwidth]{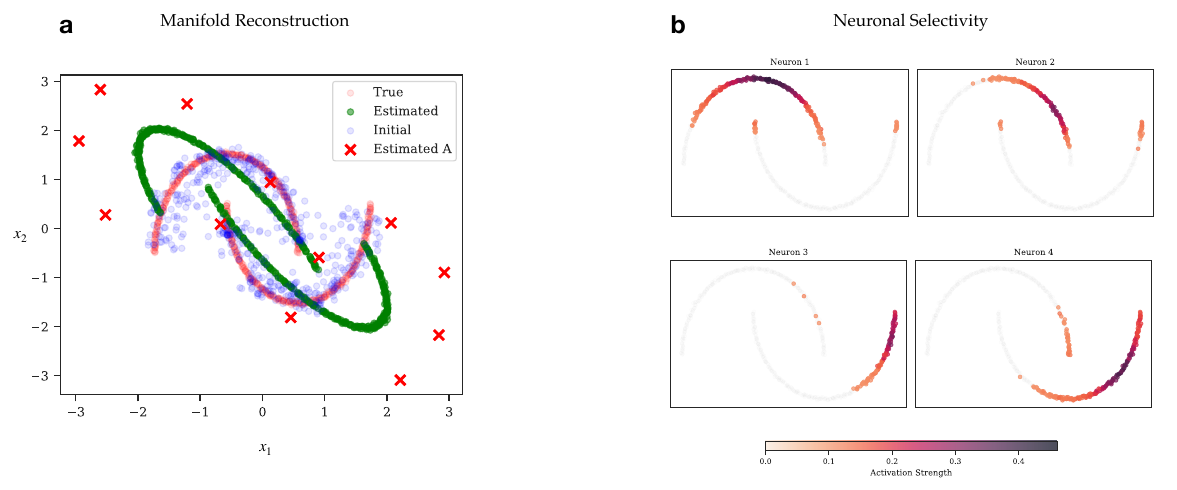}
    \caption{Applying probability simplex constraints on the latent representations. (a) Predicted inputs (green) ($\hat\mX$) from the two moons dataset alongside the true values (red) and the initial value (blue) prior to the optimization process. We see that the predicted inputs is able to capture the shape and structure of the data manifold (due to similarity alignment), however, due to the relaxed nature of the optimization problem, the prediction does not align perfectly with the input. The estimated receptive fields (indicated by a red cross), also circumvents the predicted manifold along its curvature. The final $R^2$ value achieved was 0.7. (b) Neuronal selectivity, shows the input samples that the a particular neuron is getting activated for, demonstrating local receptive nature of the neurons, albeit wider than the case in \Figref{fig:ksm-kds}.}
    \label{fig: smpc-manifold-results}
\end{figure}

In the experiments above, we observe, that the SMPC architecture is able to learn the underlying latent representations in a neurally plausible manner by incorporating bottom up and top-down interactions. Furthermore, the circuit parameters are able to capture features needed to construct a generative map as shown by an $R^2$ value which is greater than $0.5$. Nonetheless, the implicit generative maps thus estimated differ from the true values, due to the relaxed nature of the optimization problem that does not impose any constraints on the structure of the kernel in the latent space. The imposition of constraints on the kernel structure (governed by $\mOmega$) while maintaining local plasticity rules for synaptic learning will be investigated in a future work.

\section{Conclusion}
\label{sec: conclusion}



We introduce a kernel similarity matching framework for learning representations within a generative context \rev{for both overcomplete and undercomplete representations}. We explore this method in the realm of dictionary learning and show that learning the kernel function in the latent space alongside the representations can facilitate learning of generative maps between the representation and the input space respectively. Our approach formulates this hypothesis as a constrained optimization problem with the constraints applied on the latent kernel structure, extending prior works by \citep{pehlevan_why_2018,luther_kernel_2022}. To address the computational challenges inherent in the optimization of the objective, we developed a separable formulation that distributes the objective and the constraints across individual samples, thereby enabling online optimization. Finally, we develop an ADMM based algorithm to formulate the gradient dynamics of the optimization variables, and draw connections to standard optimization algorithms deployed used in similar settings (\ref{sec: computational_considerations})

The versatility of our framework is further illustrated, when we adapt it to learn a generative framework for manifold learning. Here, we reconstruct the input manifold piece-wise by learning receptive fields along its curvature. This is achieved by imposing a probability simplex constraint on the latent representations, which is inspired by metabolic limits on neuronal firing. The representations implicitly capture the geometry of the input manifold by learning receptive fields along the data manifold which in conjunction with the learned latent representations, piece-wise approximates the input data manifold. This is in stark contrast with geometry preserving feature learning techniques leveraging kernel matching in prior works\rev{\citep{sengupta_manifold-tiling_2018,luther_kernel_2022}}, where the learned representations do not capture an implicit generative map. Furthermore, in our case, the number of active neurons within the population of latent representations estimate the intrinsic dimension of the input manifold, which can lead to a normative way of understanding how geometric structure of the input data manifold play a role in shaping what the population encodes.

Building on these results, we extend our kernel similarity objective to design a predictive coding network based on the gradient dynamics of the associated variables. The proposed architecture facilitates representation learning by integrating bottom-up interactions through similarity matching and top-down interactions via predictive coding, with an inter-neuron system regulating these interactions. We applied this framework to simulated datasets and showed that the network can estimate the latent representations with reasonable reconstruction of the input data.

\rev{As mentioned before, Table \ref{tab:review-comparison-1} in \S \ref{sec:intro} summarizes the key features of our approach in comparison with other kernel-based approaches for representation learning. However, there are certain key limitations for now in the proposed approach. We demonstrated sparse coding using SMPC on simulated dataset with low mutual coherence, we found that the reconstruction quality drops as the coherence increases, with the algorithm failing to reconstruct properly under high mutual coherence in high dimensions. A similar result can be seen in \Figref{fig: smpc-manifold-results}, where we apply the manifold constraints on the latent representations in the SMPC architecture, leading to the conclusion that in the absence of structural constraints on the representation kernel, the generated input predictions — though may be similar in structure to the true input — deviate from the actual underlying values}. Addressing this limitation in future work may involve imposing explicit constraints on the latent kernel structure while preserving local plasticity to ensure biological plausibility. Another challenge the architecture faces, as depicted in \Figref{fig: smpc-architecture}(a) is the \textit{weight transport problem}, where weight of one synapse ($\mOmega$) needs to be copied to a different synapse ($\mOmega^T$). Recent work by \citep{millidge2020relaxing} provides possible approaches in dealing with this problem. Other recent arguments on neural plausibility explore three-factor Hebbian rules for learning which \citep{pogodin2020kernelized} use to learn representations in a kernelized setting by regulating the mutual information between the input and the latent representations, known as the \textit{information bottleneck principle} \citep{tishby2000information}. 

While we have focused on two specific kernel matching scenarios for representation learning, our framework’s inherent flexibility invites exploration of alternative kernel functions and the unique generative behaviors they induce. Although our proposed architecture offers a mechanistic model for synaptic plasticity, its hierarchical extensions (\Figref{fig: smpc-architecture}(b)) and implementation in physiological settings remain promising directions for future research.




\section{Acknowledgements}
We thank Dr. Abiy Tasissa, Department of Mathematics, Tufts University, Prof. Venkatesh Murthy, Harvard University,  and members of the Ba lab for their valuable feedback and discussions. SC was partially supported by the Kempner Graduate Fellowship and both SC and DB were partially supported by the National Science Foundation (grant DMS-2134157). The computations in this paper were performed on the Faculty of Arts and Sciences Research Computing (FASRC) cluster at Harvard University, and was supported in part by the Kempner Institute for the Study of Natural and Artificial Intelligence at Harvard University. This research was carried out in part thanks to funding from the Canada First Research Excellence Fund, awarded to PM through the Healthy Brains, Healthy Lives initiative at McGill University. PM acknowledges the support of the Natural Sciences and Engineering Research Council of Canada (NSERC) (grant numbers RGPIN-2025-05676 and DGECR-2025-00255), and of the Fonds de Recherche du Qu\'ebec (grant number CB-365865).

\printbibliography

@article{tasissa_k-deep_2023,
	title = {K-{Deep} {Simplex}: {Manifold} {Learning} via {Local} {Dictionaries}},
	volume = {71},
	issn = {1941-0476},
	shorttitle = {K-{Deep} {Simplex}},
	url = {https://ieeexplore.ieee.org/abstract/document/10281397},
	doi = {10.1109/TSP.2023.3322820},
	urldate = {2024-01-19},
	journal = {IEEE Transactions on Signal Processing},
	author = {Tasissa, Abiy and Tankala, Pranay and Murphy, James M. and Ba, Demba},
	year = {2023},
	note = {Conference Name: IEEE Transactions on Signal Processing},
	pages = {3741--3754},
}

@misc{chen_simple_2020,
	title = {A {Simple} {Framework} for {Contrastive} {Learning} of {Visual} {Representations}},
	url = {http://arxiv.org/abs/2002.05709},
	doi = {10.48550/arXiv.2002.05709},
	urldate = {2023-12-22},
	publisher = {arXiv},
	author = {Chen, Ting and Kornblith, Simon and Norouzi, Mohammad and Hinton, Geoffrey},
	month = jun,
	year = {2020},
	note = {arXiv:2002.05709 [cs, stat]},
}

@article{boutin_sparse_2021,
	title = {Sparse deep predictive coding captures contour integration capabilities of the early visual system},
	volume = {17},
	issn = {1553-7358},
	url = {https://dx.plos.org/10.1371/journal.pcbi.1008629},
	doi = {10.1371/journal.pcbi.1008629},
	language = {en},
	number = {1},
	urldate = {2023-12-20},
	journal = {PLOS Computational Biology},
	author = {Boutin, Victor and Franciosini, Angelo and Chavane, Frederic and Ruffier, Franck and Perrinet, Laurent},
	editor = {Latham, Peter E.},
	month = jan,
	year = {2021},
	pages = {e1008629},
}

@article{keller_predictive_2018,
	title = {Predictive {Processing}: {A} {Canonical} {Cortical} {Computation}},
	volume = {100},
	issn = {0896-6273},
	shorttitle = {Predictive {Processing}},
	url = {https://www.sciencedirect.com/science/article/pii/S0896627318308572},
	doi = {10.1016/j.neuron.2018.10.003},
	number = {2},
	urldate = {2023-12-20},
	journal = {Neuron},
	author = {Keller, Georg B. and Mrsic-Flogel, Thomas D.},
	month = oct,
	year = {2018},
	pages = {424--435},
}

@article{boyd_distributed_2011,
	title = {Distributed {Optimization} and {Statistical} {Learning} via the {Alternating} {Direction} {Method} of {Multipliers}},
	volume = {3},
	issn = {1935-8237, 1935-8245},
	url = {https://www.nowpublishers.com/article/Details/MAL-016},
	doi = {10.1561/2200000016},
	language = {English},
	number = {1},
	urldate = {2023-12-20},
	journal = {Foundations and Trends® in Machine Learning},
	author = {Boyd, Stephen and Parikh, Neal and Chu, Eric and Peleato, Borja and Eckstein, Jonathan},
	month = jul,
	year = {2011},
	note = {Publisher: Now Publishers, Inc.},
	pages = {1--122},
}

@article{luther_kernel_2022,
	title = {Kernel similarity matching with {Hebbian} networks},
	volume = {35},
	url = {https://proceedings.neurips.cc/paper_files/paper/2022/hash/0f98645119923217a245735c2c4d23f4-Abstract-Conference.html},
	language = {en},
	urldate = {2023-12-20},
	journal = {Advances in Neural Information Processing Systems},
	author = {Luther, Kyle and Seung, Sebastian},
	month = dec,
	year = {2022},
	pages = {2282--2293},
}

@misc{tolooshams_stable_2022,
	title = {Stable and {Interpretable} {Unrolled} {Dictionary} {Learning}},
	url = {http://arxiv.org/abs/2106.00058},
	doi = {10.48550/arXiv.2106.00058},
	urldate = {2023-12-20},
	publisher = {arXiv},
	author = {Tolooshams, Bahareh and Ba, Demba},
	month = aug,
	year = {2022},
	note = {arXiv:2106.00058 [cs, eess, stat]},
}

@article{olshausen_sparse_1997,
	title = {Sparse coding with an overcomplete basis set: {A} strategy employed by {V1}?},
	volume = {37},
	issn = {0042-6989},
	shorttitle = {Sparse coding with an overcomplete basis set},
	url = {https://www.sciencedirect.com/science/article/pii/S0042698997001697},
	doi = {10.1016/S0042-6989(97)00169-7},
	number = {23},
	urldate = {2023-12-20},
	journal = {Vision Research},
	author = {Olshausen, Bruno A. and Field, David J.},
	month = dec,
	year = {1997},
	pages = {3311--3325},
}

@article{bogacz_tutorial_2017,
	series = {Model-based {Cognitive} {Neuroscience}},
	title = {A tutorial on the free-energy framework for modelling perception and learning},
	volume = {76},
	issn = {0022-2496},
	url = {https://www.sciencedirect.com/science/article/pii/S0022249615000759},
	doi = {10.1016/j.jmp.2015.11.003},
	urldate = {2023-12-20},
	journal = {Journal of Mathematical Psychology},
	author = {Bogacz, Rafal},
	month = feb,
	year = {2017},
	pages = {198--211},
}

@inproceedings{sengupta_manifold-tiling_2018,
	title = {Manifold-tiling {Localized} {Receptive} {Fields} are {Optimal} in {Similarity}-preserving {Neural} {Networks}},
	volume = {31},
	url = {https://proceedings.neurips.cc/paper_files/paper/2018/hash/ee14c41e92ec5c97b54cf9b74e25bd99-Abstract.html},
	urldate = {2023-12-20},
	booktitle = {Advances in {Neural} {Information} {Processing} {Systems}},
	publisher = {Curran Associates, Inc.},
	author = {Sengupta, Anirvan and Pehlevan, Cengiz and Tepper, Mariano and Genkin, Alexander and Chklovskii, Dmitri},
	year = {2018},
}

@article{pehlevan_why_2018,
	title = {Why {Do} {Similarity} {Matching} {Objectives} {Lead} to {Hebbian}/{Anti}-{Hebbian} {Networks}?},
	volume = {30},
	issn = {0899-7667},
	url = {https://doi.org/10.1162/neco_a_01018},
	doi = {10.1162/neco_a_01018},
	number = {1},
	urldate = {2023-12-20},
	journal = {Neural Computation},
	author = {Pehlevan, Cengiz and Sengupta, Anirvan M. and Chklovskii, Dmitri B.},
	month = jan,
	year = {2018},
	pages = {84--124},
}

@article{olshausen_emergence_1996,
	title = {Emergence of simple-cell receptive field properties by learning a sparse code for natural images},
	volume = {381},
	copyright = {1996 Springer Nature Limited},
	issn = {1476-4687},
	url = {https://www.nature.com/articles/381607a0},
	doi = {10.1038/381607a0},
	language = {en},
	number = {6583},
	urldate = {2023-12-20},
	journal = {Nature},
	author = {Olshausen, Bruno A. and Field, David J.},
	month = jun,
	year = {1996},
	note = {Number: 6583
Publisher: Nature Publishing Group},
	pages = {607--609},
}

@article{rao_predictive_1999,
	title = {Predictive coding in the visual cortex: a functional interpretation of some extra-classical receptive-field effects},
	volume = {2},
	copyright = {1999 Nature America Inc.},
	issn = {1546-1726},
	shorttitle = {Predictive coding in the visual cortex},
	url = {https://www.nature.com/articles/nn0199_79},
	doi = {10.1038/4580},
	language = {en},
	number = {1},
	urldate = {2023-12-20},
	journal = {Nature Neuroscience},
	author = {Rao, Rajesh P. N. and Ballard, Dana H.},
	month = jan,
	year = {1999},
	note = {Number: 1
Publisher: Nature Publishing Group},
	pages = {79--87},
}

@inproceedings{hajinezhad2016nonnegative,
  title={Nonnegative matrix factorization using ADMM: Algorithm and convergence analysis},
  author={Hajinezhad, Davood and Chang, Tsung-Hui and Wang, Xiangfeng and Shi, Qingjiang and Hong, Mingyi},
  booktitle={2016 IEEE International Conference on Acoustics, Speech and Signal Processing (ICASSP)},
  pages={4742--4746},
  year={2016},
  organization={IEEE}
}

@article{wang2019global,
  title={Global convergence of ADMM in nonconvex nonsmooth optimization},
  author={Wang, Yu and Yin, Wotao and Zeng, Jinshan},
  journal={Journal of Scientific Computing},
  volume={78},
  pages={29--63},
  year={2019},
  publisher={Springer}
}

@article{lecun1998gradient,
  title={Gradient-based learning applied to document recognition},
  author={LeCun, Yann and Bottou, L{\'e}on and Bengio, Yoshua and Haffner, Patrick},
  journal={Proceedings of the IEEE},
  volume={86},
  number={11},
  pages={2278--2324},
  year={1998},
  publisher={Ieee}
}

@article{pedregosa2011scikit,
  title={Scikit-learn: Machine learning in Python},
  author={Pedregosa, Fabian and Varoquaux, Ga{\"e}l and Gramfort, Alexandre and Michel, Vincent and Thirion, Bertrand and Grisel, Olivier and Blondel, Mathieu and Prettenhofer, Peter and Weiss, Ron and Dubourg, Vincent and others},
  journal={Journal of machine learning research},
  volume={12},
  number={Oct},
  pages={2825--2830},
  year={2011}
}

@article{boutin2020effect,
  title={Effect of top-down connections in Hierarchical Sparse Coding},
  author={Boutin, Victor and Franciosini, Angelo and Ruffier, Franck and Perrinet, Laurent},
  journal={Neural Computation},
  volume={32},
  number={11},
  pages={2279--2309},
  year={2020},
  publisher={MIT Press One Rogers Street, Cambridge, MA 02142-1209, USA journals-info~…}
}

@article{bredies2008linear,
  title={Linear convergence of iterative soft-thresholding},
  author={Bredies, Kristian and Lorenz, Dirk A},
  journal={Journal of Fourier Analysis and Applications},
  volume={14},
  pages={813--837},
  year={2008},
  publisher={Springer}
}

@inproceedings{martins2016softmax,
  title={From softmax to sparsemax: A sparse model of attention and multi-label classification},
  author={Martins, Andre and Astudillo, Ramon},
  booktitle={International conference on machine learning},
  pages={1614--1623},
  year={2016},
  organization={PMLR}
}

@article{olshausen1996emergence,
  title={Emergence of simple-cell receptive field properties by learning a sparse code for natural images},
  author={Olshausen, Bruno A and Field, David J},
  journal={Nature},
  volume={381},
  number={6583},
  pages={607--609},
  year={1996},
  publisher={Nature Publishing Group UK London}
}

@article{tolooshams2021stable,
  title={Stable and interpretable unrolled dictionary learning},
  author={Tolooshams, Bahareh and Ba, Demba},
  journal={arXiv preprint arXiv:2106.00058},
  year={2021}
}

@article{friston2005theory,
  title={A theory of cortical responses},
  author={Friston, Karl},
  journal={Philosophical transactions of the Royal Society B: Biological sciences},
  volume={360},
  number={1456},
  pages={815--836},
  year={2005},
  publisher={The Royal Society London}
}

@article{mumford1992computational,
  title={On the computational architecture of the neocortex: II The role of cortico-cortical loops},
  author={Mumford, David},
  journal={Biological cybernetics},
  volume={66},
  number={3},
  pages={241--251},
  year={1992},
  publisher={Springer}
}

@article{millidge2020relaxing,
  title={Relaxing the constraints on predictive coding models},
  author={Millidge, Beren and Tschantz, Alexander and Seth, Anil and Buckley, Christopher L},
  journal={arXiv preprint arXiv:2010.01047},
  year={2020}
}

@article{pogodin2020kernelized,
  title={Kernelized information bottleneck leads to biologically plausible 3-factor hebbian learning in deep networks},
  author={Pogodin, Roman and Latham, Peter},
  journal={Advances in Neural Information Processing Systems},
  volume={33},
  pages={7296--7307},
  year={2020}
}

@article{tishby2000information,
  title={The information bottleneck method},
  author={Tishby, Naftali and Pereira, Fernando C and Bialek, William},
  journal={arXiv preprint physics/0004057},
  year={2000}
}

@article{bahroun2021normative,
  title={A normative and biologically plausible algorithm for independent component analysis},
  author={Bahroun, Yanis and Chklovskii, Dmitri and Sengupta, Anirvan},
  journal={Advances in Neural Information Processing Systems},
  volume={34},
  pages={7368--7384},
  year={2021}
}

@online{onlinebioplausible,
  author = {Dmitri Chklovskii},
  title = {The search for biologically plausible neural computation: The conventional approach},
  year = {2016},
  url = {https://www.offconvex.org/2016/11/03/MityaNN1/},
  urldate = {2016-11-03}
}

@article{millidge2021predictive,
  title={Predictive coding: a theoretical and experimental review},
  author={Millidge, Beren and Seth, Anil and Buckley, Christopher L},
  journal={arXiv preprint arXiv:2107.12979},
  year={2021}
}

@article{foldiak2003sparse,
  title={Sparse coding in the primate cortex},
  author={Foldiak, Peter},
  journal={The handbook of brain theory and neural networks},
  year={2003},
  publisher={MIT press}
}

@article{baddeley1996efficient,
  title={An efficient code in V1?},
  author={Baddeley, Roland},
  journal={Nature},
  volume={381},
  number={6583},
  year={1996}
}

@article{xu2016empirical,
  title={An empirical study of ADMM for nonconvex problems},
  author={Xu, Zheng and De, Soham and Figueiredo, Mario and Studer, Christoph and Goldstein, Tom},
  journal={arXiv preprint arXiv:1612.03349},
  year={2016}
}

@inproceedings{zimmermann2021contrastive,
  title={Contrastive learning inverts the data generating process},
  author={Zimmermann, Roland S and Sharma, Yash and Schneider, Steffen and Bethge, Matthias and Brendel, Wieland},
  booktitle={International Conference on Machine Learning},
  pages={12979--12990},
  year={2021},
  organization={PMLR}
}

@misc{tibshiranicnvx,
  author       = {Pasumarthi, Ramakumar and Al-Shedivat, Maruan and Tibshirani, Ryan},
  title        = {Lecture Notes in Convex Optimization},
  year         = {2016},
  url          = {https://www.stat.cmu.edu/~ryantibs/convexopt-F16/scribes/prox-grad-scribed.pdf},
  publisher = {Stanford University},
}

@article{lipshutz2022interneurons,
  title={Interneurons accelerate learning dynamics in recurrent neural networks for statistical adaptation},
  author={Lipshutz, David and Pehlevan, Cengiz and Chklovskii, Dmitri B},
  journal={arXiv preprint arXiv:2209.10634},
  year={2022}
}

@article{simoncelli2001natural,
  title={Natural image statistics and neural representation},
  author={Simoncelli, Eero P and Olshausen, Bruno A},
  journal={Annual review of neuroscience},
  volume={24},
  number={1},
  pages={1193--1216},
  year={2001},
  publisher={Annual Reviews 4139 El Camino Way, PO Box 10139, Palo Alto, CA 94303-0139, USA}
}

@article{van2008visualizing,
  title={Visualizing data using t-SNE.},
  author={Van der Maaten, Laurens and Hinton, Geoffrey},
  journal={Journal of machine learning research},
  volume={9},
  number={11},
  year={2008}
}

@article{carandini2012normalization,
  title={Normalization as a canonical neural computation},
  author={Carandini, Matteo and Heeger, David J},
  journal={Nature reviews neuroscience},
  volume={13},
  number={1},
  pages={51--62},
  year={2012},
  publisher={Nature Publishing Group UK London}
}

@article{chalk2017sensory,
  title={Sensory noise predicts divisive reshaping of receptive fields},
  author={Chalk, Matthew and Masset, Paul and Deneve, Sophie and Gutkin, Boris},
  journal={PLoS Computational Biology},
  volume={13},
  number={6},
  pages={e1005582},
  year={2017},
  publisher={Public Library of Science San Francisco, CA USA}
}

@article{chung2021neural,
  title={Neural population geometry: An approach for understanding biological and artificial neural networks},
  author={Chung, SueYeon and Abbott, Larry F},
  journal={Current opinion in neurobiology},
  volume={70},
  pages={137--144},
  year={2021},
  publisher={Elsevier}
}

@article{jazayeri2021interpreting,
  title={Interpreting neural computations by examining intrinsic and embedding dimensionality of neural activity},
  author={Jazayeri, Mehrdad and Ostojic, Srdjan},
  journal={Current opinion in neurobiology},
  volume={70},
  pages={113--120},
  year={2021},
  publisher={Elsevier}
}

@article{zhang2024introduction,
  title={An Introduction to Bilevel Optimization: Foundations and applications in signal processing and machine learning},
  author={Zhang, Yihua and Khanduri, Prashant and Tsaknakis, Ioannis and Yao, Yuguang and Hong, Mingyi and Liu, Sijia},
  journal={IEEE Signal Processing Magazine},
  volume={41},
  number={1},
  pages={38--59},
  year={2024},
  publisher={IEEE}
}

@article{ghodsi2006dimensionality,
  title={Dimensionality reduction a short tutorial},
  author={Ghodsi, Ali},
  journal={Department of Statistics and Actuarial Science, Univ. of Waterloo, Ontario, Canada},
  volume={37},
  number={38},
  pages={2006},
  year={2006}
}

@article{kreutz2003dictionary,
  title={Dictionary learning algorithms for sparse representation},
  author={Kreutz-Delgado, Kenneth and Murray, Joseph F and Rao, Bhaskar D and Engan, Kjersti and Lee, Te-Won and Sejnowski, Terrence J},
  journal={Neural computation},
  volume={15},
  number={2},
  pages={349--396},
  year={2003},
  publisher={MIT Press One Rogers Street, Cambridge, MA 02142-1209, USA journals-info~…}
}

@article{rao2024sensory,
  title={A sensory--motor theory of the neocortex},
  author={Rao, Rajesh PN},
  journal={Nature Neuroscience},
  volume={27},
  number={7},
  pages={1221--1235},
  year={2024},
  publisher={Nature Publishing Group US New York}
}

@article{deneve2016circular,
  title={Circular inference: mistaken belief, misplaced trust},
  author={Deneve, Sophie and Jardri, Renaud},
  journal={Current Opinion in Behavioral Sciences},
  volume={11},
  pages={40--48},
  year={2016},
  publisher={Elsevier}
}

@article{jiang2021improved,
  title={Improved training of sparse coding variational autoencoder via weight normalization},
  author={Jiang, Linxing Preston and de la Iglesia, Luciano},
  journal={arXiv preprint arXiv:2101.09453},
  year={2021}
}

@article{tolooshams2024interpretable,
  title={Interpretable deep learning for deconvolutional analysis of neural signals},
  author={Tolooshams, Bahareh and Matias, Sara and Wu, Hao and Temereanca, Simona and Uchida, Naoshige and Murthy, Venkatesh N and Masset, Paul and Ba, Demba},
  journal={bioRxiv},
  year={2024},
  publisher={Cold Spring Harbor Laboratory Preprints}
}

@incollection{dempe2020bilevel,
  title={Bilevel optimization},
  author={Dempe, Stephan and Zemkoho, Alain},
  booktitle={Springer optimization and its applications},
  volume={161},
  year={2020},
  publisher={Springer}
}

@article{mayzel2023homeostatic,
  title={Homeostatic synaptic normalization optimizes learning in network models of neural population codes},
  author={Mayzel, Jonathan and Schneidman, Elad},
  journal={bioRxiv},
  pages={2023--03},
  year={2023},
  publisher={Cold Spring Harbor Laboratory}
}

@article{turrigiano2004homeostatic,
  title={Homeostatic plasticity in the developing nervous system},
  author={Turrigiano, Gina G and Nelson, Sacha B},
  journal={Nature reviews neuroscience},
  volume={5},
  number={2},
  pages={97--107},
  year={2004},
  publisher={Nature Publishing Group UK London}
}

@article{lee2006efficient,
  title={Efficient sparse coding algorithms},
  author={Lee, Honglak and Battle, Alexis and Raina, Rajat and Ng, Andrew},
  journal={Advances in neural information processing systems},
  volume={19},
  year={2006}
}

@misc{o1978hippocampus,
  title={The hippocampus as a cognitive map},
  author={O’Keefe, J},
  year={1978},
  publisher={Oxford University Press}
}

@article{chaudhuri2019intrinsic,
  title={The intrinsic attractor manifold and population dynamics of a canonical cognitive circuit across waking and sleep},
  author={Chaudhuri, Rishidev and Ger{\c{c}}ek, Berk and Pandey, Biraj and Peyrache, Adrien and Fiete, Ila},
  journal={Nature neuroscience},
  volume={22},
  number={9},
  pages={1512--1520},
  year={2019},
  publisher={Nature Publishing Group US New York}
}

@article{knudsen1978center,
  title={Center-surround organization of auditory receptive fields in the owl},
  author={Knudsen, Eric I/ and Konishi, Masakazu},
  journal={Science},
  volume={202},
  number={4369},
  pages={778--780},
  year={1978},
  publisher={American Association for the Advancement of Science}
}

@article{da2013tuning,
  title={Tuning in to sound: frequency-selective attentional filter in human primary auditory cortex},
  author={Da Costa, Sandra and van der Zwaag, Wietske and Miller, Lee M and Clarke, Stephanie and Saenz, Melissa},
  journal={Journal of Neuroscience},
  volume={33},
  number={5},
  pages={1858--1863},
  year={2013},
  publisher={Soc Neuroscience}
}

@article{hubel1959receptive,
  title={Receptive fields of single neurones in the cat’s striate cortex},
  author={Hubel, David H and Wiesel, Torsten N and others},
  journal={J physiol},
  volume={148},
  number={3},
  pages={574--591},
  year={1959}
}

@book{borg2007modern,
  title={Modern multidimensional scaling: Theory and applications},
  author={Borg, Ingwer and Groenen, Patrick JF},
  year={2007},
  publisher={Springer Science \& Business Media}
}

@article{donoho2006compressed,
  title={Compressed sensing},
  author={Donoho, David L},
  journal={IEEE Transactions on information theory},
  volume={52},
  number={4},
  pages={1289--1306},
  year={2006},
  publisher={IEEE}
}

@article{wang2013projection,
  title={Projection onto the probability simplex: An efficient algorithm with a simple proof, and an application},
  author={Wang, Weiran and Carreira-Perpin{\'a}n, Miguel A},
  journal={arXiv preprint arXiv:1309.1541},
  year={2013}
}

@article{bach2012optimization,
  title={Optimization with sparsity-inducing penalties},
  author={Bach, Francis and Jenatton, Rodolphe and Mairal, Julien and Obozinski, Guillaume and others},
  journal={Foundations and Trends{\textregistered} in Machine Learning},
  volume={4},
  number={1},
  pages={1--106},
  year={2012},
  publisher={Now Publishers, Inc.}
}

@article{daubechies2004iterative,
  title={An iterative thresholding algorithm for linear inverse problems with a sparsity constraint},
  author={Daubechies, Ingrid and Defrise, Michel and De Mol, Christine},
  journal={Communications on Pure and Applied Mathematics: A Journal Issued by the Courant Institute of Mathematical Sciences},
  volume={57},
  number={11},
  pages={1413--1457},
  year={2004},
  publisher={Wiley Online Library}
}

@article{ba2020deeply,
  title={Deeply-Sparse Signal rePresentations (DS$^2$P)},
  author={Ba, Demba},
  journal={IEEE Transactions on Signal Processing},
  volume={68},
  pages={4727--4742},
  year={2020},
  publisher={IEEE}
}

@article{choi2004kernel,
  title={Kernel isomap},
  author={Choi, Heeyoul and Choi, Seungjin},
  journal={Electronics letters},
  volume={40},
  number={25},
  pages={1612--1613},
  year={2004},
  publisher={IET}
}

@book{cox2000multidimensional,
  title={Multidimensional scaling},
  author={Cox, Trevor F and Cox, Michael AA},
  year={2000},
  publisher={CRC press}
}

@article{sanchez2021gentle,
  title={A gentle introduction to graph neural networks},
  author={Sanchez-Lengeling, Benjamin and Reif, Emily and Pearce, Adam and Wiltschko, Alexander B},
  journal={Distill},
  volume={6},
  number={9},
  pages={e33},
  year={2021}
}

@article{lipshutz2023normative,
  title={Normative framework for deriving neural networks with multicompartmental neurons and non-hebbian plasticity},
  author={Lipshutz, David and Bahroun, Yanis and Golkar, Siavash and Sengupta, Anirvan M and Chklovskii, Dmitri B},
  journal={PRX Life},
  volume={1},
  number={1},
  pages={013008},
  year={2023},
  publisher={APS}
}

@inproceedings{minden2018biologically,
  title={Biologically plausible online principal component analysis without recurrent neural dynamics},
  author={Minden, Victor and Pehlevan, Cengiz and Chklovskii, Dmitri B},
  booktitle={2018 52nd Asilomar Conference on Signals, Systems, and Computers},
  pages={104--111},
  year={2018},
  organization={IEEE}
}

@article{vidal2005generalized,
  title={Generalized principal component analysis (GPCA)},
  author={Vidal, Rene and Ma, Yi and Sastry, Shankar},
  journal={IEEE transactions on pattern analysis and machine intelligence},
  volume={27},
  number={12},
  pages={1945--1959},
  year={2005},
  publisher={IEEE}
}

@inproceedings{pehlevan2014hebbian,
  title={A Hebbian/anti-Hebbian network derived from online non-negative matrix factorization can cluster and discover sparse features},
  author={Pehlevan, Cengiz and Chklovskii, Dmitri B},
  booktitle={2014 48th Asilomar Conference on Signals, Systems and Computers},
  pages={769--775},
  year={2014},
  organization={IEEE}
}

@article{donoho2003optimally,
  title={Optimally sparse representation in general (nonorthogonal) dictionaries via l1 minimization},
  author={Donoho, David L and Elad, Michael},
  journal={Proceedings of the National Academy of Sciences},
  volume={100},
  number={5},
  pages={2197--2202},
  year={2003},
  publisher={The National Academy of Sciences}
}
\newpage
\begin{appendices}

\section{Proof of Lemma \ref{lem: Min-Min Lemma}}

\label{app: min-min-lemma}

\begin{proof}
  Let us assume without loss of generality, $\underset{\vx}{\min}\,\underset{\vy}{\min}\,\,f(\vx, \vy) > \underset{\vy}{\min}\,\underset{\vx}{\min}\,\,f(\vx, \vy)$.\\
  Let $f(\vx^*, \vy^*) = \underset{\vx}{\min}\,\underset{\vy}{\min}\,\,f(\vx, \vy)$ and $(\vx^\dag, \vy^\dag) = \underset{\vy}{\min}\,\underset{\vx}{\min}\,\,f(\vx, \vy)$\\
  then, we have $f(\vx^*, \vy^*) > f(\vx^\dag, \vy^\dag)$. However,
  \begin{align*}
      f(\vx^\dag, \vy^\dag) & \ge \underset{\vx}{\min}\,\,f(\vx, \vy^\dag)\\
      & \ge \underset{\vx}{\min}\,\underset{\vy}{\min}\,\,f(\vx, \vy)\\
      & = f(\vx^*, \vy^*) \implies \text{A contradiction}
  \end{align*}
  Therefore, the only possibility is $f(\vx^*, \vy^*) = f(\vx^\dag, \vy^\dag)$ or $\underset{\vx}{\min}\,\underset{\vy}{\min}\,\,f(\vx, \vy) = \underset{\vy}{\min}\,\underset{\vx}{\min}\,\,f(\vx, \vy)$
\end{proof}



\subsection{Proof of proposition \ref{prop: ksm-obj}}
\label{app: prop-ksm-obj}
\begin{proof}
    Starting from $P_1$, using lemma \ref{lem: Min-Min Lemma}, we can rewrite $P_1$ as:
    \begin{align*}
        \underset{\mZ, \mA}{\min}\,\,\frac{1}{2T}\|\mX - \mA\mZ\|^2_F + \frac{\lambda}{T}\|\mZ\|_1 + \frac{\omega}{2} \|\mA\|_F^2 \qquad \omega > 0\\
    \end{align*}
    where, we have changed the order of minimization. The inner optimization is now a strongly convex problem with respect to the dictionary $\mA$ and has a unique optimum at $\mA^* = \dfrac{\mX\mZ}{T}\left(\dfrac{\mZ\mZ^T}{T} + \omega I\right)^{-1}$. Plugging it back into the objective and defining $\mH = \dfrac{\mZ\mZ^T}{T} + \omega I$, we end up with the objective in $Q_1$.
\end{proof}


\subsection{Proof of proposition \ref{prop: ksm-manifold}}
\label{app: prop-ksm-manifold}
\begin{proof}
    Similar to before we switch the order of optimization in $P_2$ using  lemma \ref{lem: Min-Min Lemma} to obtain:
    \begin{align*}
        \underset{\mZ, \mA}{\min}\,\,\frac{1}{2T}\|\mX - \mA\mZ\|^2_F + \frac{\omega}{T}\sum\limits_{k=1}^K\sum\limits_{l=1}^T z_{kl}\|\vx_l - \va_k\|_2^2 \qquad \omega > 0\\
    \end{align*}
    Solving the inner optimization w.r.t. $\mA$ we end up with $\mA^* = (1+\omega)\dfrac{\mX\mZ^T\mH^{-1}}{T}$, where $\mH = \dfrac{\mZ\mZ^T}{T} + \omega \mD,\,\,\text{and}\,\,\mD = \text{Diag}\left( \dfrac{\mZ\mathbf{1}_T}{T} \right)$, is a diagonal matrix. Substituting in back in $P_2$ we get
    \begin{equation}
        \label{eq: ksm-manifold-scaled}
        \begin{aligned}
           \underset{\mZ, \mH}{\min}\,\,& -\frac{(1+\omega)^2}{2T^2}\Tr(\mX^T\mX\mZ^T\mH^{-1}\mZ)\\
             \text{s.t.}\,\,&\,\, \mZ^T\mathbf{1}_K = \mathbf{1}_T\\
             &\,\,\mZ \ge 0\\
             &\,\,\mH = \frac{\mZ\mZ^T}{T} + \omega\mD 
        \end{aligned}
     \end{equation}
    Scaling the objective in \eqref{eq: ksm-manifold-scaled} by $1 / (1+\omega)^2$ we get the objective in $Q_2$.
\end{proof}

\newpage

\section{Projection onto the probability simplex induces subtractive normalization}
\label{app: proj-simplex}
In this section we show that how the projection onto the probability simplex induces a normalization effect on the overall population activity. We consider the following optimization problem depicting the output of the simplex projection:

\begin{equation}
    \begin{aligned}
        \vz = \arg\undermin{x\in\sR^D}\,\,&\frac{1}{2}\|\vx - \vy\|^2\\
        \text{s.t.}&\,\,\vx^T\mathbf{1} = 1\\
        &\,\,\vx \ge 0
    \end{aligned}
\end{equation}
where, $\vz$ denotes the population activity and $\vy$ denotes the input to the population. Without loss of generality, we assume the input activity to be sorted such that $y_1 \ge y_2 \ge \dots y_D$. The solution of the above optimization problem is given by:
\begin{equation}
    \label{eq: proj-simplex}
    \begin{aligned}
        z_i = \max\left(0, y_i - \lambda (\vy)\right)
    \end{aligned}
\end{equation}
with $\lambda(\vy) = \dfrac{1}{\rho}\left(\sum\limits_{i=1}^\rho y_i - 1\right)$, where $\rho = \max\{1\le i \le D: y_i - \dfrac{1}{\rho}\left(\sum\limits_{i=1}^\rho y_i - 1\right) > 0\}$. The threshold $\lambda(\vy)$ is adaptive and is derived by computing a running mean of the population activity sorted in decreasing order (if the activity is not sorted, then we sort the activity first and then compute the sum). We refer the reader to \cite{wang2013projection} for a detailed proof of the above result.

We see from \Eqref{eq: proj-simplex} that the output of the simplex projection involves a thresholding operation on the input activity with an adaptive threshold computed by calculating a sort of running mean over the population activity which is then subtracted from the input. We call this operation \textit{subtractive normalization}, as the result of this operation is that the output activity is normalized to sum to unity.


\newpage

\section{ADMM dynamics for solving the KSM problem}
\label{app: ksm-dynamics}
In this section we layout the gradient dynamics for each step of the ADMM optimization articulated in Algorithm \ref{algo:admm_sep} for the separable Lagrangian of the kernel similarity matching problem with sparsity constraints on the latent space.

\subsection{Dynamics with Sparsity constraints on the latent space}
\label{app: subsec-ksm-sc-dynamics}
Given objective:

\begin{equation}
    \begin{aligned}
    \undermin{\mW, \vz_i, \mP_i}&\,\,-\frac{1}{T}\sum\limits_{i=1}^T \left(\vx_i^T \mW^T \vz_i\right) + \frac{1}{2T}\sum\limits_{i=1}^T \Tr\left( \mW^T\mP_i\mW \right) + \frac{\lambda}{T}\sum\limits_{i=1}^T\|\vz_i\|_{1}\\
    \text{s.t.}&\,\,\mP_i = \vz_i\vz_i^T+ \omega \mathbf{I} \qquad \forall\,\,i \in \{1, 2, \dots, T\}
    \end{aligned}
\end{equation}
The above objective involves optimizing over $\mW, \vz_i, \mP_i$. For a given iteration $k$ of the ADMM algorithm (Algorithm \ref{algo:admm_sep}), the update dynamics of each of the optimization subproblem in each step of the ADMM algorithm is given below, where $\tau$ represents the current iteration in the optimization subproblem.

\begin{itemize}
    \item \underline{Z-Step}\\
    
     The Z-Step for the $(k+1)^{th}$ epoch involves the following optimization:
     $$\vz_i^{k+1} \gets \underset{\vz}{\arg\min}\,\,\Ls_i(\vz_i, \mP_i^{k},\mM_i^k)$$ 
     where,
     $$\Ls_i = -\vx_i^T\mW\vz_i + \frac{1}{2}\Tr(\mW^T\mP_i\mW) + \lambda\|\vz_i\|_1 + \rho\left\|\vz_i\vz_i^T + \omega \mI - \mP_i + \frac{\mM_i}{\rho}\right\|_F^2$$
    The Lagrangian described before is a \textit{quartic} function in $\vz_i$ involving smooth and non-smooth functions, so we resort to gradient based methods to optimize for $\vz_i$. Specifically, we use proximal gradient descent methods to account for the $\ell_1$ norm regularization. The update dynamics for the $\tau+1^{th}$ iteration of the proximal method is given as:
	 \begin{equation}
        \label{eq: z-step}
		\begin{aligned}
			\vz_i^{(\tau+1), k} &= \gS_{\frac{\lambda}{\eta}}\left[\vz_i^{(\tau), k} - \eta_Z\left\{-\mW^{k}\vx_i + 2\rho \left(\vz_i^{(\tau), k}\vz_i^{{(\tau), k}^T} + \omega\mI -\mP_i^{k} +\frac{\mM_i^{k}}{\rho}\right)\vz_i^{(\tau), k}  \right\}\right]\\
            \vz_i^{k+1}&= \vz_i^{(J), k}
		\end{aligned}
	\end{equation}
    where we run the \textit{Z-Step} for $J$ iterations of proximal gradient descent.

    \item \underline{P-Step}\\
    
    The gradient dynamics for the update to $\mP_i$ is given as:
	\begin{equation}
		\begin{aligned}
			\mP_i^{(\tau+1), k+1} = \mP_i^{(\tau), k+1} + \eta_P \left[-\frac{\mW^{k}\mW^{k\,\,T}}{2} +  \rho \left(  \vz_i^{k+1}\vz_i^{{k+1}\,T} + \omega \mI - \mP_i^{(\tau), k+1} + \frac{\mM_i^{k}}{\rho}\right)  \right]
		\end{aligned}
	\end{equation}
	In our experiments, we kept $\eta_P \gg \eta_Z, \eta_W$, essentially using fixed point update for $\mP_i$, given as:
    $$\mP_i^{k+1} = \frac{1}{\rho}\left( \mM_i^{k} -\frac{\mW^{k}\mW^{k\,\,T}}{2}\right) +  \vz_i^{k+1}\vz_i^{{k+1}^T} + \omega \mI $$

    \item \underline{M-Step}\\
    
    The updates for the Lagrange multiplier $\mM_i$ is given as:
	\begin{equation}
		\begin{aligned}
			\mM_i^{k+1} = \mM_i^{k} +  \rho \left(\vz_i^{k+1}\vz_i^{{k+1}\,T} + \omega \mI - \mP_i^{k+1}\right)
		\end{aligned}
	\end{equation} 

    \item \underline{W-Step}\\
    
    The gradient dynamics for the $\mW$ update is given as:
	\begin{equation}
		\begin{aligned}
			\mW^{(\tau+1), k+1} = \mW^{(\tau), k+1} + \frac{\eta_W}{T} \sum\limits_{i=1}^{N_b}\left( \vz_i^{k+1}\vx_i^T  - \mP_i^{k}\mW^{(\tau), k+1}  \right)
		\end{aligned}
	\end{equation}
\end{itemize}

\subsubsection{Equivalence to Schedule 1 to ISTA (proof of theorem 1)}
\label{app: subsubsec-equi-ISTA}
\begin{proof}
Appendix \ref{app: subsec-ksm-sc-dynamics} outlines the proximal gradient dynamics for the \textit{Z-Step} under \textit{Schedule 1}. Since every gradient step in the \textit{Z-Step} is followed by a \textit{P-Step}, then for the $(\tau)^{th}$ iteration of the \textit{Z-Step}, the update dynamics for $\vz_i$ can be written as:

\begin{equation}
    \begin{aligned}
        \vz_i^{(\tau), k} &= \gS_{\frac{\lambda}{\eta}}\left[\vz_i^{(\tau-1), k} - \eta_Z\left\{-\mW^{k-1}\vx_i + 2\rho \left(\vz_i^{(\tau-1), k}\vz_i^{{(\tau-1), k}^T} + \omega\mI -\mP_i^{(\tau-1),k} +\frac{\mM_i^{k-1}}{\rho}\right)\vz_i^{(\tau-1), k}  \right\}\right]
    \end{aligned}
\end{equation}

with the corresponding \textit{P-Step} given as:
\begin{equation}
    \begin{aligned}
    \label{eq: p-step-sched-1-proof}
            \mP_i^{(\tau), k} &= \frac{1}{\rho}\left( \mM_i^{k-1} -\frac{\mW^{k-1}\mW^{(k-1)\,\,T}}{2}\right) +  \vz_i^{(\tau), k}\vz_i^{{(\tau), k}^T} + \omega \mI
    \end{aligned}
\end{equation}
Using the result of the \textit{P-Step} in \Eqref{eq: p-step-sched-1-proof}, we can write the update for the $(\tau+1)^{th}$ iteration of the \textit{Z-Step} can be written as:
\begin{equation}
    \label{eq: z-step-sched-1-proof}
    \begin{aligned}
        \vz_i^{(\tau+1), k} &= \gS_{\frac{\lambda}{\eta}}\left[\vz_i^{(\tau), k} - \eta_Z\left\{-\mW^{k-1}\vx_i + 2\rho \left(\vz_i^{(\tau), k}\vz_i^{{(\tau), k}^T} + \omega\mI -\mP_i^{(\tau),k} +\frac{\mM_i^{k-1}}{\rho}\right)\vz_i^{(\tau), k}  \right\}\right]\\
        &= \gS_{\frac{\lambda}{\eta}}\left[\vz_i^{(\tau), k} - \eta_Z\left\{-\mW^{k-1}\vx_i + 2\rho\cdot \frac{\mW^{(k-1)}\mW^{{(k-1)}^T}}{2\rho}\vz_i^{(\tau-1), k}  \right\}\right]\\
        &= \gS_{\frac{\lambda}{\eta}}\left[ -\mW^{(k-1)}\vx_i + (\mI - \eta_\mZ\mW^{(k-1)}\mW^{(k-1)^T})\vz_i^{(\tau), k}\right]
    \end{aligned}
\end{equation}
which is the Iterative Soft Thresholding Algorithm (ISTA) for sparse coding \cite{bredies2008linear} defined by the optimization problem in \Eqref{eq:orig_sparse_coding} where the dictionary $\mA$ is given by $\mW^T$. The \textit{W-Step} then implements the dictionary update for the $k^{th}$ iteration of the algorithm to get $\mW^{k}$. Finally, we set $\mM_i^{k} = \frac{\mW^{k}\mW^{k^T}}{2}$ to complete the \textit{M-Step} under \textit{Schedule 1}. This final step, allows us to incorporate the changes from the \textit{W-Step} in the \textit{M-Step} in the current iteration, thereby ensuring the constraint on $\mP_i$ is satisfied. This completes the proof.

\end{proof}

\textbf{Note on Schedule 2:} The key difference in \textit{Schedule 2} when compared to \textit{Schedule 1} is in the \textit{M-Step} which involves setting $\mM_i^k = \mM_i^{k-1} + \rho\left(\vz_i^k\vz_i^{k^T} + \omega\mI - \mP_i^{k}\right)$. Since the \textit{Z-Step} and \textit{P-step} are performed as per \textit{Schedule 1}, we can substitute \Eqref{eq: p-step-sched-1-proof} in \textit{M-Step} definition for \textit{Schedule 2} to obtain,
\begin{equation}
    \begin{aligned}
        \mM_i^k &= \mM_i^{k-1} + \rho\left[\frac{1}{\rho}\left( \frac{\mW^{(k-1)}\mW^{(k-1)^T}}{2} - \mM_i^{k-1}\right] \right)\\
        &= \frac{\mW^{(k-1)}\mW^{(k-1)^T}}{2}
    \end{aligned}
\end{equation}
This is similar to the \textit{M-Step} in \textit{Schedule 1}, except the updates to $\mM_i$ are delayed by one iteration. While the \textit{Z-Step} for the $k+1^{th}$ iteration, follows the same dynamics as outlined in \Eqref{eq: z-step-sched-1-proof}, the \textit{P-step} updates would involve $\mM_i^{k}$, a function of $\mW^{k-1}$, resulting in the constraints on $\mP_i$ not being immediately satisfied as long as \textit{W-Step} does not reach convergence. Notably, as the \textit{Z-Step} dynamics remain unchanged alongside the \textit{W-Step}, \textit{Schedule 2} still performs ISTA based alternate minimization like \textit{Schedule 1}, which is reflected in the rates of optimization in \figref{fig: schedule-comparison}(c).

\subsection{Dynamics for simplex constraint - Manifold Learning}
\label{app: subsec-manifold}
As described in the main text, the objective for the manifold learning objective follows the same structure as the sparse dictionary learning with variation in the constraints on the latent variables. Consequently, the gradient dynamics mimic that of sparse dictionary learning with the primary difference being in the \textit{Z-Step}. For our experiments we use the scaled version of the objective as described in \Eqref{eq: ksm-manifold-scaled} as it accelerated the optimization process. The separable version of the scaled objective in \Eqref{eq: ksm-manifold-scaled} is given by and can be determined along the lines of Proposition \ref{prop: online-obj-ksm-sc}:
\begin{equation}
    \begin{aligned}
        \undermin{\mW, \vz_i, \mP_i}&\,\,-\frac{1+\omega}{T}\sum\limits_{i=1}^T \left(\vx_i^T \mW^T \vz_i\right) + \frac{1}{2T}\sum\limits_{i=1}^T \Tr\left( \mW^T\mP_i\mW \right)\\
        \text{s.t.}&\,\,\sum\limits_{j=1}^K z_{ij} = 1, \quad z_{ij} \ge 0 \qquad \forall\,\,i \in \{1, 2, \dots, T\},\quad j \in \{1, 2, \dots, K\}\\
        &\,\,\mP_i = \vz_i\vz_i^T+ \omega \mD_i\quad \text{where}\quad \mD_i = \text{diag}(\vz_i) \qquad \forall\,\,i \in \{1, 2, \dots, T\}
    \end{aligned}
\end{equation}

Consequently, the dynamics of the optimization variables are given as follows:
\begin{itemize}
    \item \underline{Z-Step}:\\
    \begin{equation}
        \begin{aligned}
            \vz_i^{(\tau+1), k+1} = \gP_S&\left[\vz_i^{(\tau), k+1} - \eta_Z\left\{-(1+\omega)\mW^{k}\vx_i + 2\rho\left(\vz_i^{(\tau), k+1}\vz_i^{{(\tau), k+1}^T} + \omega\mD_i -\mP_i^{k} +\frac{\mM_i^{k}}{\rho}\right)\vz_i^{(\tau), k+1}\right.\right.\\
            &  \left.\left. +\rho\omega\,\text{diag}\left( \vz_i^{(\tau), k+1}\vz_i^{{(\tau), k+1}^T} + \omega\mD_i  -\mP_i^{k} +\frac{\mM_i^{k}}{\rho}\right)\right\}\right]
        \end{aligned}
    \end{equation}
    where, $\gP_S$ is the simplex projection operator (\cite{martins2016softmax}).

    \item \underline{P-Step}:\\
    
    \begin{equation}
        \begin{aligned}
            \mP_i^{(\tau+1), k+1} = \mP_i^{(\tau), k+1} + \eta_P \left[-\frac{\mW^{k}\mW^{k\,\,T}}{2} +  \rho \left(  \vz_i^{k+1}\vz_i^{{k+1}\,T} + \omega \mD_i - \mP_i^{(\tau), k+1} + \frac{\mM_i^{k}}{\rho}\right)  \right]
        \end{aligned}
    \end{equation}
    Like before, we set $\eta_P \gg \eta_Z, \eta_W$ and use the fixed point update for $\mP_i$.

    \item \underline{M-Step}:\\
    
    \begin{equation}
        \begin{aligned}
            \mM_i^{k+1} = \mM_i^{k} +  \rho \left(\vz_i^{k+1}\vz_i^{{k+1}\,T} + \omega \mD_i - \mP_i^{k+1}\right)
        \end{aligned}
    \end{equation}

    \item \underline{W-Step}:\\
    
    \begin{equation}
        \begin{aligned}
            \mW^{(\tau+1), k+1} = \mW^{(\tau), k+1} + \frac{\eta_W}{T} \sum\limits_{i=1}^{N_b}\left( (1+\omega)\vz_i^{k+1}\vx_i^T  - \mP_i^{k}\mW^{(\tau), k+1}  \right)
        \end{aligned}
    \end{equation}
\end{itemize}



\newpage

\section{Saddle Point property of the SMPC objective}

Here we describe the saddle point property that allows us to go from proposition \ref{prop: smpc-online} to proposition \ref{prop: smpc-minmax}. Our derivation builds upon the one described in \cite{pehlevan_why_2018}, Appendix A. From Theorem 3 in that paper , if $f : \sR^n \times \sR^m \to \sR$ satisfies the saddle point property then, $\exists \va^* \in \sR^n, \vb^* \in \sR^m$, such that $\forall \va \in \sR^n, \vb \in \sR^m$:
\begin{equation*}
    \begin{aligned}
        f(\va^*, \vb) &\le f(\va^*, \vb^*) \le f(\va, \vb^*)
    \end{aligned}
\end{equation*}
then,
\begin{align*}
    \undermax{\vb}\, \undermin{\va} f(\va, \vb) = \undermin{\va}\, \undermax{\vb} f(\va, \vb) = f(\va^*, \vb^*)
\end{align*}

To show that the objective in \Eqref{eq:smpc} satisfies the saddle point property, we define $f$ as follows:
\begin{equation}
\label{eq: f-saddle}
f(\mY, \mM, \mA) = -\frac{1}{T} \Tr(\mA^T\mY) + \frac{1}{2T}\Tr(\mY^T\mM\mY) - \frac{1}{4}\Tr(\mM^T\mM) + \frac{1}{T}g(\mY)
\end{equation}

where, $g(\mY) = \undermin{\mZ\in\gC}\,\,h(\mZ, \mY)$. For the sparse coding case, $h(\mY, \mZ) =  \frac{\rho}{2} \|\mY - \mOmega\mZ\|_F^2 + \lambda\|\mZ\|_{1, 1}$ and $h(\mY, \mZ) = \frac{\rho}{2} \|\mY - \mOmega\mZ\|_F^2$ for the case with probability simplex constraints (where $\gC = \Delta^K$, $\Delta^K$ denoting the probability simplex in $K$ dimensions). Since $g$ is a strongly convex function in $\mZ$ (over a closed and non empty set), the minimum of $g$ w.r.t. $\mZ$ is unique and exists, in spite of it being analytically intractable. Our definition of $f$ is motivated by the fact that the order of optimization between the variables $\mM$ and $\mZ$ can easily be swapped as the objective defined in proposition \ref{prop: smpc-online} is completely separable in $\mM$ and $\mZ$. 

Let $\mY^*$ and $\mM^*$ define the saddle point for the function $f$ obtained by setting $\nabla f = 0$ w.r.t. to both $\mY$ and $\mM$, such that $\mY^*$ and $\mM^*$ satisfy:

\begin{equation}
    \label{eq: saddle-point-soln}
    \begin{aligned}
        \mM^* &= \frac{1}{T}\mY^*\mY^{*^T}\\
        \mM^*\mY^* + \nabla g(\mY^*) &= \mA
    \end{aligned}
\end{equation}

To see if the saddle point property for \Eqref{eq: f-saddle} holds, we examine the inequalities outlined in the definition above. To check the first inequality, we compute
\begin{align*}
    & f(\mY^*, \mM^*, \mA) - f(\mY^*, \mM, \mA) \\
    &= \frac{1}{2T}\Tr\left(\mY^{*^T}(\mM^* - \mM)\mY^*\right) - \frac{1}{4}\Tr(\mM^{*^2} - \mM^2)\\
    &= \frac{1}{2}\Tr(\mM^*(\mM^* - \mM)) - \frac{1}{4}\Tr(\mM^{*^2} - \mM^2) \quad (\text{\Eqref{eq: saddle-point-soln}})\\
    &= \frac{1}{4}\|\mM - \mM^*\|_F^2 \ge 0
\end{align*}

For the second inequality, we compute
\begin{align*}
    & f(\mY, \mM^*, \mA) - f(\mY^*, \mM^*, \mA) \\
    &= -\frac{1}{T}\Tr\left(\mA^T(\mY - \mY^*)\right) + \frac{1}{2T}\Tr(\mY^T\mM^*\mY) - \frac{1}{2T}\Tr(\mY^{*^T}\mM^*\mY^*) + \frac{1}{T} \left(g(\mY) - g(\mY^*)\right)\\
    &= -\frac{1}{T}\Tr\left(\left[\mY^{*^T}\mM^* + \nabla g(\mY^*)^T\right](\mY-\mY^*)\right) +\frac{1}{2T}\Tr(\mY^T\mM^*\mY) - \frac{1}{2T}\Tr(\mY^{*^T}\mM^*\mY^*) + \frac{1}{T} \left(g(\mY) - g(\mY^*)\right) \quad (\text{\Eqref{eq: saddle-point-soln}})\\
    &= \frac{1}{2T}\Tr\left((\mY^* - \mY)^T\mM^*(\mY^* - \mY)\right)  + \frac{1}{T} \left(g(\mY) - g(\mY^*) - \nabla g(\mY^*)^T(\mY - \mY^*)\right)
\end{align*}

Since $\mM^*$ is a positive semidefinite matrix (\Eqref{eq: saddle-point-soln}), the first term on the RHS in the above expression is non-negative. For the second term, we note that the function $h(\mY, \mZ)$ is convex in both $\mY$ and $\mZ$. Then $g(\mY) = \undermin{\mZ}\,\,h(\mY, \mZ)$ is also convex in $\mY$. To see this, consider for any $t \in [0, 1]$ and some $\mY_1, \mY_2$, we can write:
\begin{align*}
    g(t\mY_1 + (1-t)\mY_2) &= \undermin{\mZ}\,\,h(t\mY_1 + (1-t)\mY_2, \mZ)\\
    &\le\,\, h(t\mY_1 + (1-t)\mY_2, t\mZ_1 + (1-t)\mZ_2)\quad(\text{for some arbitrary } \mZ_1, \mZ_2)\\
    &\le\,\, t h(\mY_1, \mZ_1) + (1-t)h(\mY_2, \mZ_2)\\
\end{align*}
Since the choice of $\mZ_1, \mZ_2$ was arbitrary, we can choose $\mZ_1 = \arg\undermin{Z}\,\,h(\mY_1, \mZ)$ and $\mZ_2 = \arg\undermin{Z}\,\,h(\mY_2, \mZ)$. This implies that
\begin{align*}
    g(t\mY_1 + (1-t)\mY_2) &\le t g(\mY_1) + (1-t)g(\mY_2)
\end{align*}
thereby proving that $g$ is convex. Consequently for any convex function $g$, we have
\begin{align*}
    g(\mY) \ge g(\mY^*) + \nabla g(\mY^*)^T(\mY - \mY^*)
\end{align*}
for all $\mY, \mY^*$. This implies that the second term on the RHS of the second inequality is non-negative, thereby proving that the objective in \Eqref{eq: f-saddle} satisfies the saddle point property.



\newpage
\section{Deriving the SMPC architecture}
\label{app: smpc}
Here we derive the \rev{discretized} gradient dynamics for the SMPC objective outlined in Proposition \ref{prop: smpc-minmax} and \rev{Algorithm \ref{algo:smpc-bioplausible}} which plays a key role in formulating the circuit diagram for the SMPC objective. We focus on sparsity constraints on the latents with similar arguments being extended to the case with probability simplex constraints.
The objective for the SMPC model is described as follows in Proposition \ref{prop: smpc-minmax}:
\begin{align}
    \underset{\mOmega, \mW}{\min}\,\undermax{\mM}\,\undermin{\mY, \mZ}&\,\frac{1}{T}\sum\limits_{i=1}^N \left(-\vx_i^T\mW^T\vy_i + \frac{1}{2T}\Tr(\mW^T\mW) + \lambda\|\vz_i\|_1 + \frac{\rho}{2}\|\vy_i - \mOmega\vz_i\|^2 + \frac{1}{2}\vy_i^T\mM\vy_i - \frac{1}{4}\Tr(\mM^T\mM)\right)
\end{align}
To construct the circuit given by \Figref{fig: smpc-architecture}, we first formulate the gradient dynamics of the objective above, and then use it to come up with a predictive coding network (\cite{bogacz_tutorial_2017}). We define $\Delta \va^k = \va^{k+1} - \va^k$ as the update to the variable $\va$ at the $k^{th}$ iteration. Further, we define $\bm{\zeta}_i = \rho(\mOmega\vz_i-y_i)$ as the interneuron activity for the $i^{th}$ sample, tracking the error between the encoding $\vy_i$ and prediction from top-down interactions with the latent $\vz_i$. The resulting update dynamics for each of the optimization variables can then be articulated as:

\begin{align}
    \label{eq: y-update-smpc}
    \underline{\text{Y-update}} \quad \Delta \vy_i^k &= \eta_\mY\left(\mW\vx_i - \mM\vy_i^k + \bm{\zeta}_i^k\right)
\end{align}

\begin{equation}
    \begin{aligned}
        \label{eq: z-update-smpc}
        \underline{\text{Z-update}} \quad \Delta\vv_i^{k} &= -\mOmega^T\bm{\zeta}_i^k + \frac{1}{\eta_\mZ}(\vz_i^k - \vv_i^k)\\
        \vz_i^{k+1} &= f(\vv_i^{k+1}, \vz_i^k)
    \end{aligned}
\end{equation}

\begin{equation}
    \label{eq: zeta-update-smpc}
    \begin{aligned}
        \underline{\bm\zeta\text{-update}} \quad \Delta\zeta_i^{k} &= \mOmega\vz_i^{k+1} - \vy_i^{k+1} - \rho^{-1}\zeta^k\\
    \end{aligned}
\end{equation}
\begin{equation}
    \label{eq: synaptic-update-smpc}
    \begin{aligned}
        \underline{\mOmega\text{-update}} \quad \Delta\mOmega^k &= -\eta_\mOmega\left[\frac{1}{T}\sum\limits_{i=1}^T \left(\bm{\zeta}_i^{k+1}\vz_i^{k+1^T}\right)\right]\\
        \underline{\text{W-update}} \quad \Delta\mW^k &=   \eta_\mW\left[\frac{1}{T}\left(\sum\limits_{i=1}^N\vy_i^{k+1}\vx_i^T - \mW^{k}\right)\right]\\
        \underline{\text{M-update}} \quad  \Delta\mM^k &= \eta_\mM\left[\frac{1}{T}\left(\sum\limits_{i=1}^N\vy_i^{k+1}\vy_i^{k+1^T} - \mM^k\right)\right]
    \end{aligned}
\end{equation}
Here, $f(\vv_i^{k+1}, \vz_i^k) = \vv_i^{k+1} - \eta_\mZ\lambda \text{sign}(\vz_i^k)$ (where $\lambda \text{sign}(\vz_i^k)$ comes from the subgradient of the $L_1$ norm) for the sparsity constraint. For the probability simplex constraint,  $f(\vv_i^{k+1}, \vz_i^k) = \max (0, \vv_i^{k+1} - \lambda(\vv_i^{k+1}))$, where $\lambda(\vv_i^{k+1})$ is calculated as described in Appendix \ref{app: proj-simplex}.

\Eqref{eq: y-update-smpc}, \Eqref{eq: z-update-smpc}, \Eqref{eq: zeta-update-smpc} model the neural dynamics whereas \Eqref{eq: synaptic-update-smpc} models the synaptic dynamics. The dynamics presented above allows us to draw the predictive coding network shown in \Figref{fig: smpc-architecture}. Under stochastic gradient update paradigm, \Eqref{eq: synaptic-update-smpc} would involve single sample approximations for the gradients resulting in an online learning rule which results in slower and noisy convergence. For our simulations, we used batch gradient methods to accelerate the convergence of the model.




\subsection{Initialization and Implementation Details for the SMPC model}
In this section we discuss the initialization and the training implementation for the SMPC architecture. 

\begin{itemize}
    \item \textbf{2D dataset} For the simple 2D dataset, we sample the final latents, $\mZ$, independently from $\gN(0, 1)$. To initialize $\mOmega$, we first computed $\mH = \frac{\mZ\mZ^T}{T} + \omega \mI$, with $\omega=0.1$ and $T$ denoting the total number of samples. Thereafter, we computed the Cholesky decomposition to get a lower triangular matrix which we inverted and transposed to initialize $\mOmega$. The idea being $\mOmega^T\mOmega$ should resemble $\mH^{-1}$. We then initialized $\mY = \mOmega \mZ$. The elements of $\mW$ were sampled independently from $\gN(0, 1)$ and $\mM$ was initialized to identity matrix. The total training was run for 6000 iterations.
    
    \item \textbf{High Dimensional Synthetic Data} For the high dimensional simulated data, we initialized the representations $\mZ = \mZ_{\text{true}} + \texttt{noise}$ where the elements of the noise term were sampled independently from $\gN(0, 1)$. We chose a relaxed initialization of $\mOmega$, by allowing the elements to be sampled independently from $\gN(0, 1)$. $\mY, \mW, \mM$ was initialized the same as before. The total training was run for 10000 iterations.

    \item \textbf{Moons dataset with simplex constraints} Similar to the previous step, we initialized the representations $\mZ = \mZ_{\text{true}} + \texttt{noise}$ where the elements of the noise term were sampled independently from $\gN(0, \sigma^2)$, where $\sigma = 5.0$. To obtain $\mZ_{\text{true}}$, we ran alternate minimization as outlined in \cite{tasissa_k-deep_2023} with a value of $\omega = 0.1$ (\Eqref{eq:manifold_learning}) for $500$ iterations. We initialized $\mOmega$ similar to the 2D dataset, by computing $\mH=\dfrac{\mZ\mZ^T}{T} + \omega \mD$ where $\mD = diag(\dfrac{\mZ\bm{1}_T}{T})$, with $T$ denoting the number of examples. $\mY, \mW, \mM$ was initialized the same as before. The total training was run for 8000 iterations.
\end{itemize}

We implemented the optimization process, using PyTorch autograd function to compute the gradients. The interneuron values were set to their fixed points and the batch size was set to the dataset length to speed up the training process.


\newpage

\section{p-sparsity measure}
\label{app: p-sparsity}
The p-sparsity measure is a metric to compute the effective sparsity of a given vector. Let $\vx \in \mathbb{R}^n$ be a vector, then the p-sparsity measure ($S_p(\vx)$) is defined as:
\begin{equation}
    S_p(\vx) =  \left( \frac{1}{N_p}\right)\left[\frac{\left( \frac{1}{n-1}\right)\left(\sum\limits_{i=1}^n|x_i - \bar{x}|^p\right)}{\left(\frac{1}{n}\right)\sum\limits_{i=1}^n |x_i|^p} \right]^{\frac{1}{p}}
\end{equation}
where, $N_p = \left( \dfrac{(n-1)^{p-1} + 1}{n^{p-1}} \right)^{\frac{1}{p}}$ and $\bar{x} = \dfrac{1}{n}\sum\limits_{i=1}^n x_i$, where $\vx = [x_1, x_2, \dots, x_n]^T$. For the case when $p=1$, we get $N_p = 1$ and the p-sparsity measure is given by:
\begin{equation}
    S_1(\vx) = \frac{\sum\limits_{i=1}^n |x_i - \bar{x}|}{\sum\limits_{i=1}^n |x_i|}
\end{equation}
When $\|\vx\|_0 = 1$, then $S_1(\vx) = 1$, i.e. the given vector $\vx$ is highly sparse. Similarly, $S_1(x)$ is closer to 0 when $\vx$ is less sparse.
\end{appendices}






\end{document}